\documentclass[reprint,pre]{revtex4-1}
\usepackage{graphicx}
\usepackage{epstopdf}
\usepackage{epsfig}
\usepackage{amssymb,amsmath,stmaryrd,tabularx}
\usepackage{wrapfig}
\usepackage{bm}
\usepackage[center]{subfigure}
\usepackage{enumerate}
\usepackage{hyperref}
\usepackage{tikz}

\usepackage{forest}

\renewcommand{\v}[1]{\ensuremath{\mathbf{#1}}} 
\let\crossprod=\times
\renewcommand{\times}{\cdot} 
\newcommand{\gv}[1]{\ensuremath{\mbox{\boldmath$ #1 $}}} 
\newcommand{\diff}{\mathrm{d}}
\newcommand{\pd}[2]{\frac{\partial #1}{\partial #2}}

\newcommand{\grad}[1]{\gv{\nabla} #1} 
\renewcommand{\div}[1]{\gv{\nabla} \cdot #1} 

\newcommand{\laplacian}[1]{\grad^2 #1}

\newcommand{\pdt}[1]{\partial_t #1}

\newcommand{\Aes}{{A_{\bf es}}}
\newcommand{\Ade}{{A_{\bf de}}}
\newcommand{\Ads}{{A_{\bf ds}}}

\newcommand{\Vold}{{\Omega_{\bf d}}}

\begin{document}
\title{Controlling wetting with electrolytic solutions: phase-field simulations of a droplet-conductor system}

\newcommand{\NBI}{Niels Bohr Institute, University of Copenhagen, Blegdamsvej 17, DK-2100 Copenhagen, Denmark.}
\author{Gaute Linga}
\email{linga@nbi.dk}
\affiliation{\NBI}
\author{Asger J.~S.~Bolet}
\affiliation{\NBI}
\author{Joachim Mathiesen}
\affiliation{\NBI}


\begin{abstract}
  The wetting properties of immiscible two-phase systems are crucial in a wide range of applications, from lab-on-a-chip devices to field-scale oil recovery.
  It has long been known that effective wetting properties can be altered by the application of an electric field; a phenomenon coined as electrowetting.
  Here, we consider theoretically and numerically a single droplet sitting on an (insulated) conductor, i.e., within a capacitor.
  The droplet consists of a pure phase without solutes, while the surrounding fluid contains a symmetric monovalent electrolyte, and the interface between them is impermeable.
  Using nonlinear Poisson--Boltzmann theory, we present a theoretical prediction of the dependency of the  apparent contact angle on the applied electric potential.
  We then present well-resolved dynamic simulations of electrowetting using a phase-field model, where the entire two-phase electrokinetic problem, including the electric double layers (EDLs), is resolved.
  The simulations show that, while the contact angle on scales smaller than the EDL is unaffected by the application of an electric field, an apparent contact angle forms on scales beyond the EDL.
  This contact angle relaxes in time towards a saturated apparent contact angle.
  The dependency of the contact angle upon applied electric potential is in good compliance with the theoretical prediction.
  The only phenomenological parameter in the prediction is shown to only depend on the permeability ratio between the two phases.
  Based on the resulting unified description, we obtain an effective expression of the contact angle which can be used in more macroscopic numerical simulations, i.e.~where the electrokinetic problem is not fully resolved.
\end{abstract}

\maketitle

\section{Introduction}
\label{sec:introduction}
Precisely controlling the effective wetting properties of droplets in immiscible two-phase flows is desirable in many applications, from fabricating microfluidic devices \cite{pollack2002,srinivasan2004,lee2000} and electronic displays \cite{beni1981,beni1981b,beni1982,hayes2003} to understanding the microscopic dynamics of enhanced oil recovery, which has field-scale consequences \cite{hassenkam2011,hilner2015,rezaeidoust2009,pedersen2016,hiorth2010}.
Lippmann already in the 19th century \cite{lippmann1875,mugele2005} laid the groundwork for the field of \emph{electrowetting}, by making the observation that applying an electric field indeed \emph{can} change the wetting behaviour of conductive liquid-liquid systems.
The depencence of the contact angle $\theta$ on the applied electric potential $V_0$ could be described by a quadratic law,
\begin{equation}
  \cos \theta = \cos \theta_0 + \frac{1}{2} B V_0^2,
  \label{eq:lippmann}
\end{equation}
where $\theta_0$ is the contact angle in the absence of electrical fields, and $B$ is a phenomenological parameter.
Eq.~\eqref{eq:lippmann} can also be inferred from Gibbs' adsorption isotherm \cite{monroe2006,levich1962}.

Theoretical and experimental works have explained the basic mechanisms of electrowetting, particularly in the case of conducting liquids \cite{mugele2005,mugele2009b}.
Careful experiments show that the contact angle described by Eq.~\eqref{eq:lippmann} is a macroscopic effect, apparent only on scales beyond the insulator thickness \cite{mugele2007}.
Two notable remaining open issues within electrowetting are (1) the dynamics of the contact line \cite{eck2009}, and (2) the effect of electrolytes in either of the phases on the wetting \cite{mugele2009b}.

The latter point was explored theoretically by \citet{monroe2006,monroe2006b}, who considered interfaces between two immiscible electrolytic solutions (ITIES), and obtained a transcendental expression for the contact angle of a droplet sitting on an isolated, grounded plate using an energy minimization approach.
In contrast to ``conventional'' electrowetting systems, the phases in ITIES systems contain ions which cannot pass over to the other phase (nor the plate).
The apparent contact angle in the case of conductive liquids can only become more acute with the application of a potential, while the latter work showed that contact angles in the presence of electrolytes (and in the absence of flow) could become both obtuse or acute depending on the concentrations, permittivities and applied potential.

For conductive liquids with low net concentration of charge, the leaky-dielectric model is admissible.
Originally, this model was proposed by \citet{taylor1966} (and revisited by \citet{melcher1969}) to describe the distortion of drops in electric fields. 
Since advection and diffusion of charges is neglected in this model, the electric double layers (EDLs), characterized by the Debye length, are not resolved.
As shown rigorously by \citet{schnitzer2015}, it can be seen as a thin Debye layer limit of the full electro\emph{kinetic} model \cite{zholkovskij2002,saville1997}.
However, when ionic effects are important and charges are not constrained to the liquid-liquid interface, the more detailed level of description (i.e., resolving the full electrokinetic model) is necessary.
Several authors have considered the full model in the absence of boundaries (i.e.~for droplets immersed in a liquid). 
\citet{berry2013} presented a sharp-interface combined level-set/volume-of-fluid method to simulate such systems, as an enhancement compared to the leaky-dielectric simulations by \citet{tomar2007} and the charge conservative model by \citet{lopez-herrera2011}.
\Citet{eck2009} provided the first direct simulation studies of dynamic electrowetting with electrolytes.
The model used in the latter work belongs to the leaky-dielectric type, as the mobility does not depend on concentration.
However, it contains a concentration regularisation parameter which introduces a length scale, and effectively sets the thickness of the Debye layer.
A similar model and a more detailed study was carried out by \citet{nochetto2014}.
Other works have adopted a more macroscopic viewpoint and used the electrowetting contact angle as an input to model effective behaviour on the microfluidic scale \cite{lu2007,walker2006,walker2009}.
On the other hand, there are a number of assumptions underpinning the purely theoretical works of \citet{monroe2006}, and as experiments remain sparse, simulations would be of interest to test validity of, and extensions to, the theory.
To the authors' knowledge, there has been no systematic numerical study of the direct dependency of the contact angle on applied electric potential for a fully resolved electrohydrodynamic model with partially soluble electrolytes.

In this work, we consider theoretically and numerically the effect of an applied potential on the wetting properties of an immiscible two-phase system consisting of a single droplet placed on an insulated electrode.
The droplet phase is non-conducting, while the surrounding fluid contains an electrolyte, and all interfaces are taken to be impermeable.
Such approximations are valid for many industrially and geologically relevant systems such as oil-in-water flows \cite{berry2013}.
Using Poisson--Boltzmann theory and following the approach of Ref.~\cite{monroe2006}, we develop a theoretical prediction for the apparent contact angle dependency on applied potential.
In our simulations, we use the thermodynamically consistent and frame-invariant model for two-phase electrokinetic flow which was proposed by \citet{campillo-funollet2012}.
This phase-field model combines the Nernst--Planck equation for chemical transport, the Poisson equation for electrostatics, the Cahn--Hilliard equation for the description of the interface, and the Navier--Stokes equations for fluid flow.
Using a recently introduced solver \cite{linga2018b} for this model, we simulate electrowetting dynamically.
We demonstrate explicitly that the contact angle is only apparent on scales beyond the Debye length, whereas the \emph{microscopic} contact angle remains unaffected.
Our main finding is that the apparent contact angle dependency is well described by the theoretical prediction, in particular when our only phenomenological quantity, the effective screening area, is modelled as a function solely of the ratio between the permittivities.
This, microscopically viewed, \emph{apparent} contact angle, can thus be turned into a \emph{fixed} contact angle boundary condition which can be used for simulations on more macroscopic scales.


\section{Model system}
\label{sec:model}

\begin{figure}[htb]
  \includegraphics[width=0.99\columnwidth]{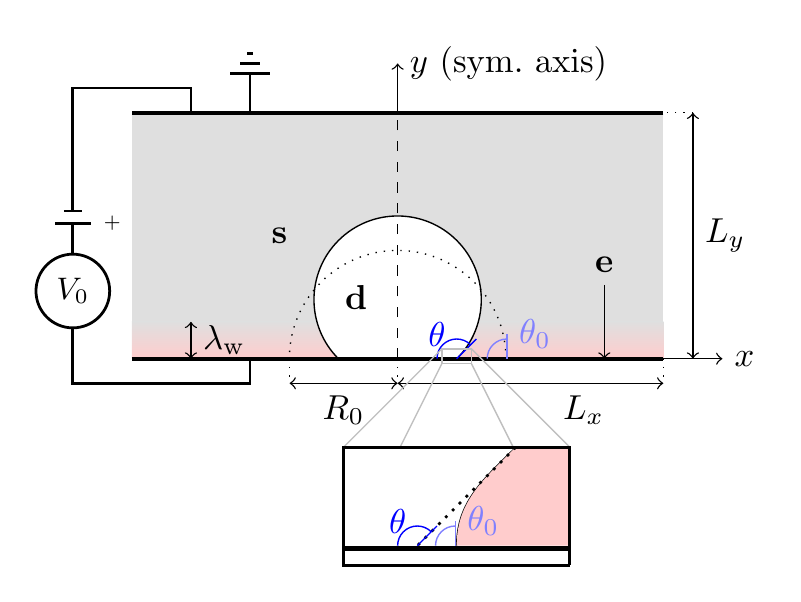}
  \caption{\label{fig:schematic}
    Schematic set-up of the numerical experiment.
    Here, \textbf{d} indicates the droplet phase, \textbf{s} indicates the surrounding phase, and \textbf{e} indicates the electrode.
    The figure shows the final state after the application of a potential difference $V_0$ between the two electrodes.
    Due to the dissolved electrolytes in phase \textbf{s}, an electric double layer, characterized by the Debye length $\lambda_{\rm s}$ is formed near the lower electrode, and an \emph{apparent} contact angle $\theta$ is formed.
    Also indicated with a dotted line is the initial state of the droplet (where $V_0=0$), forming the contact angle $\theta_0$.
    Note that the simulations considered herein exploit the indicated axial symmetry of the problem.
    A close-up view of the contact line shows how the contact angle $\theta_0$ persists on small scales, whereas the apparent contact angle $\theta$ is only evident on sufficiently large scales.
  }
\end{figure}

We consider a droplet (phase \textbf{d}), surrounded by another fluid (phase \textbf{s}), sitting on an electrode (phase \textbf{e}) in the presence of an electric field.
A sketch of the system set-up is shown in Fig.~\ref{fig:schematic}.
Within the surrounding fluid, a binary salt is dissolved.
We denote the concentrations of these ionic species by $c_\pm$.
We consider symmetric ions, such that $z_\pm = \pm z$ are the valencies of the ions.
The ions are not allowed to pass through the liquid-liquid interface (\textbf{ds}), and the droplet contains no ions.
This set-up is representative of most oil-in-water systems and most microfluidic applications.

The substrate is held at a constant electric potential $V=V_0$, while the system is grounded far from the droplet.
We take the lower boundary, representing the electrode, to be impermeable for ions and the fluid phase, and hence assume a no-slip condition.
This assumption, which implies zero conduction through the system, is the main distinction from most of the existing literature \cite{mugele2005}.

Conversely, the top boundary mimics a reservoir and thus assumes constant concentrations, i.e.\ $c_\pm = c_0$.
Due to the impermeable boundary, an EDL is formed near the electrode, as quantified by the Debye length $\lambda_{\rm s}$ indicated in Fig.~\ref{fig:schematic}.
It is well known that the local contact angle $\theta_0$ is given by the interfacial energies between the three phases, while on scales beyond $\lambda_{\rm s}$, an apparent contact line $\theta$ is formed.
Using the presented set-up, we shall in the forthcoming consider how this apparent contact angle depends on the applied potential $V_0$.

\section{Theory}
\label{sec:theory}
Two-phase electrokinetic fluid dynamics is described by the coupled problem of solute transport, fluid flow and electrostatics.
The Nernst--Planck equation governs the chemical chemical transport,
\begin{equation}
  \pd {c_\pm} {t} + \v u \cdot \grad c_\pm = \div \left( D_\pm \grad c_\pm \mp \frac{z q_e c_\pm}{k_{\rm B} T} \v E \right),
  \label{eq:cont_NP}
\end{equation}
where $t$ is the time, $\v u$ is the fluid velocity, $\v E = - \grad V$ is the electric field, 
$D_\pm$ are the diffusivities of the ``$\pm$'' ions, $k_{\rm B}$ is Boltzmann's constant, $T$ is the temperature and $q_e$ is the elementary charge.
Electrostatic equilibrium is determined by the Poisson equation,
\begin{equation}
  \div \left( \epsilon_0 \epsilon_{\rm r} \v E \right) = \rho_e,
  \label{eq:cont_P}
\end{equation}
where $\epsilon_0$ is the vacuum permittivity, $\epsilon_{\textrm{r}}$ is the relative permittivity, and the total charge is given by $\rho_e = q_e z (c_+ - c_-)$.
The fluid flow is governed by the Navier--Stokes equations,
\begin{gather}
  \rho \left( \pdt \v u + \v u \cdot \grad \v u \right) - \mu \grad^2 \v u + \grad p = - \rho_e \grad V, \label{eq:cont_NS1}\\
  \div \v u = 0, \label{eq:cont_NS2}
\end{gather}
where $\rho$ is the density, $\mu$ is the dynamic viscosity, and $p$ is the pressure.
The equations are closed by boundary conditions and the continuity of the normal stress across the interface between the phases,
\begin{equation}
  \Big[ 2 \mu \mathcal D \v u  - p' \v I + 
  \sigma \kappa \v I + \epsilon_0 \epsilon_{\rm r} \v E \otimes \v E - \frac{1}{2} \epsilon_0 \epsilon_{\rm r} \v E^2 \v I \Big] \cdot \hat{\v n} = \v 0.
  \label{eq:cont_interface}
\end{equation}
Here, the pressure $p'$ has been redefined to absorb an osmotic contribution, $\mathcal D \v u = (\grad \v u + \grad \v u^T)/2$ is the (symmetric) strain-rate tensor, $\kappa$ is the interface curvature, and $\hat{\v n}$ is an interface normal.

\subsection{Scaled variables}
\label{sec:scaled}
We employ a standard electrokinetic scaling to obtain dimensionless variables which are more practical to work with in the following.
To this end, we introduce the dimensionless variables indicated by a tilde; such that $\tilde t = t / t^*$, $\tilde \rho = \rho/\rho^*$, $\tilde {\v u} = \v u / u^*$, $\tilde p = p/p^*$, $\tilde\mu = \mu/\mu^*$, $\tilde c = c / c^*$, $\tilde V = V/V^*$, $\tilde D_\pm = D_\pm / D^*$, $\tilde \epsilon = \epsilon_{\rm r} / \epsilon^*$, and $\tilde \sigma = \sigma / \sigma^*$.
Here, all the quantities marked by an asterisk are reference values.
Further, all length variables are scaled by a droplet reference linear size $R^*$, i.e., $\tilde x = x / R^*$.
In particular, the electric potential $V$ is scaled by the thermal voltage,
\begin{equation}
  V^* = V_T = \frac{k_{\rm B} T}{z q_e}.
\end{equation}
The remaining reference quantities are given by,
\begin{gather}
  t^* = \frac{R^*}{u^*}, \quad
  \rho^* = \frac{z q_e c^* V_T}{(u^*)^2}, \\
  D^* = u^* R^*, \quad
  p^* = z q_e c^* V_T, \quad
  \mu^* = \frac{z q_e c^* V_T R^*}{u^*}, \\
  \epsilon^* = \frac{ z q_e c^* (R^*)^2}{\epsilon_0 V_T}, \quad
  \sigma^* = z q_e c^* V_T R^* .
\end{gather}
Note that time $\tilde t$ is given in advective time units.
Adopting the chosen scaling, and subsequently skipping the tildes, now results in a model consisting of the set of equations \eqref{eq:cont_NP} 
to \eqref{eq:cont_interface}, but where $zq_e = k_B T = \epsilon_0 = 1$ and $\epsilon_{\rm r} \to \epsilon$.
For simplicity of notation we shall thus retain this normalization throughout the paper.

\subsection{Equilibrium free energy}
We are here interested in the time-asymptotic steady state of the droplet.
Since there is an impermeable no-slip boundary at $y=0$, and hence no charge transport through the system in the steady-state, the steady state will be without fluid circulation.
We can thus safely neglect the velocity field in seeking the time-asymptotic state.

We denote the phasic quantities of the concentrations by $c_i$, the (dynamic) viscosity by $\mu_i$, the permittivities by $\epsilon_i$, for phases $i \in \{ \textbf{d}, \textbf{s} \}$, and the interface energies by $\sigma_j$, for $j \in \{ \textbf{ds}, \textbf{de}, \textbf{es} \}$.
The droplet and surrounding subvolumes are denoted by $\Omega_{\bf d}$ and $\Omega_{\bf s}$, respectively.

Following \citet{monroe2006}, we write the Gibbs energy $G$ of the system as
\begin{multline}
  G = - \frac{1}{2} \sum_{i=\textbf{d},\textbf{s}} \int_{\Omega_j} \epsilon_{j} \v E^2 \, \diff \Omega \\
  + \sum_{j=\pm} \int_{\Omega_\textbf{s}} \left[ (\log {c_j} - 1) c_j + \frac{z_j}{z} c_j V \right] \diff \Omega
  + \sum_{i=\textbf{d},\textbf{s}} p \, \Omega_i \\
  + \Ads \sigma_{\textbf{ds}} + \Ade \sigma_{\textbf{de}} + \Aes \sigma_{\textbf{es}}.
  \label{eq:full_G}
\end{multline}
Here, $\Ade$ is the area between the droplet and the electrode, $\Ads$ is the area between the droplet and the surroundings, and $\Aes$ is the area between the electrode and the surrounding fluid.
Like $\Omega_{\bf s}$ and $\Aes$, this energy scales with the size of the domain, and we need to fix it by defining some reference.
The reference state can be chosen as the state without a droplet, $G_0$.
We denote the deviation from this reference by $\Delta G = G-G_0$.

In contrast to \citet{monroe2006}, we consider here a droplet which does not contain electrolytes.
Neglecting the energetic contribution of the electric field within the droplet and the charge distribution around the droplet, the deviation in Gibbs free energy from a reference state without a droplet, can in the large droplet approximation of non-linear Poisson--Boltzmann theory \cite{monroe2006} be written as
\begin{multline}
  \frac{\Delta G}{\sigma_{\textbf{ds}}} = \Ade \Bigg[ \frac{8 \sqrt{2 \epsilon_{\mathbf{s}} c_0 }}{\sigma_{\textbf{ds}}} \sinh^2\left( \frac{V_0}{4} \right)
  - \cos \theta_0 \Bigg] \\
+ \Ads + \frac{\Vold}{\sigma_{\textbf{ds}}} \Delta p .
  \label{eq:DG_1}
\end{multline}
Here, $\Delta p$ is the pressure difference across the interface, which here is to be  considered as a Lagrange multiplier.
Since Eq.~\eqref{eq:DG_1} was derived without accounting for the energy within the droplet, this expression provides an upper bound for the energy.
This can be realized by considering the contribution from the thin screening layer outside the droplet (interface \textbf{ds}) and the negative sign of the electric field inside the droplet.

\subsection{A scaling ansatz}
To somewhat simplify, we define the quantity
\begin{align}
  f_0 &= \frac{8 \sqrt{2 \epsilon_{\bf s} c_0 }}{\sigma_{\textbf{ds}}},
  \label{eq:f0}
\end{align}
which, along with the applied potential $V_0$, is predicted to be a control parameter of the system.
To incorporate the effect of screening the electric field due to the droplet, we heuristically generalize Eq.~\eqref{eq:DG_1}.
Since the electric flux into the droplet is roughly proportional to the contact area $\Ade$, we postulate that the effect can effectively be incorporated by making the modification
\begin{equation}
  \frac{\Delta G}{\sigma_{\bf ds}} = \Ade \left[ f \sinh^2\left( \frac{V_0}{4} \right) - \cos \theta_0 \right]
 + \Ads + \frac{\Vold}{\sigma_{\bf ds}} \Delta p ,
\label{eq:DG}
\end{equation}
where $f \to f_0$ in the limit of no electrical flux through the droplet (and hence no screening around). 
Note, that to be consistent with the ``upper bound'' observation made above, we must have $f \leq f_0$ for all sets of parameters.
Further, making the ansatz that $f/f_0$ should depend only on quantities present in both phases, that \emph{further} contribute to the energy in the equilibrium state (cf.~Eq.~\eqref{eq:full_G}), we have
\begin{equation}
  f = f_0 \cdot h\left( \frac{\epsilon_{\textbf{d}}}{\epsilon_{\textbf{s}}} \right),
  \label{eq:def_f}
\end{equation}
where $h \leq 1$ is an unknown function.

\subsection{Expression for the contact angle}
When the surface tension $\sigma_{\bf ds}$ is sufficiently high, and considering a two-dimensional system, we may take the droplet to be a circular cap.
We can write down expressions for the interfacial areas and the droplet volume in terms of circle radius $r$ and angle $\theta$:
\begin{gather*}
  \Ade = 2 r \sin \theta, \quad
  \Ads = 2 r \theta, \\
  \Vold = r^2 \left( \theta - \frac{1}{2} \sin 2\theta \right).
\end{gather*}
The latter yields
\begin{equation}
  r = \frac{\sqrt{\Vold}}{\sqrt{\theta - \frac{1}{2} \sin 2 \theta}}.
\end{equation}
Now, Eq.~\eqref{eq:DG} can be written as
\begin{equation}
  \frac{\Delta G}{\sigma_{\bf ds}} = 2 \Vold^{1/2} \frac{ \xi \sin \theta + \theta }{\sqrt{\theta - \frac{1}{2} \sin 2 \theta}} + \frac{\Vold}{\sigma_{\bf ds}} \Delta p .
\end{equation}
where
\begin{equation}
  \xi = f \sinh^2\left( \frac{V_0}{4} \right) - \cos \theta_0 .
\end{equation}
We need to minimize $\Delta G$ with respect to the apparent contact angle $\theta$; this amounts to finding the $\theta$ that minimizes
\begin{equation}
  \chi(\theta) = \frac{ \xi \sin \theta + \theta }{\sqrt{\theta - \frac{1}{2} \sin 2 \theta}}
\end{equation}
i.e.~solving,
\begin{multline}
  \chi'(\theta) = \\
\frac{ \xi \cos \theta + 1 }{\sqrt{\theta - \frac{1}{2} \sin 2 \theta}}
  - \frac{ \left(\xi \sin \theta + \theta \right) \left( 1 - \cos 2 \theta \right) }{ 2 \left( {\theta - \frac{1}{2} \sin 2 \theta} \right)^{3/2}} = 0.
\end{multline}
This gives
\begin{equation}
  \left( \xi + \cos \theta \right) \left( \theta \cos \theta - \sin \theta \right)  = 0 .
\end{equation}
The second factor on the left hand side is nonzero for $\theta \in (0, \pi)$.
Hence, the apparent contact angle is given by $\cos \theta = - \xi$ (which can also be verified to correspond to a minimum in $\chi$).
This can be written as
\begin{equation}
  \cos \theta = \cos \theta_0 - f \sinh^2\left( \frac{V_0}{4} \right).
  \label{eq:contact_angle_prediction}
\end{equation}
Thus we have a simple expression for what to expect from numerical simulations.

Notably, since we know from before that $f \leq f_0$, we thus have a prediction of a \emph{lower bound} for the contact angle, namely
\begin{equation}
  \cos \theta - \cos \theta_0 \geq - f_0 \sinh^2\left( \frac{V_0}{4} \right).
  \label{eq:contact_angle_prediction_ineq}
\end{equation}
Furthermore, expression \eqref{eq:contact_angle_prediction} is consistent with Lippmann's expression \eqref{eq:lippmann} in the limit of $V_0 \ll 1$.
This leads us to the identification
\begin{equation}
  B = - \frac{f}{8} ,
\end{equation}
and hence we have obtained a prediction of the phenomenological parameter $B$.

We shall check the validity of Eqs.~\eqref{eq:contact_angle_prediction} and \eqref{eq:def_f} numerically in the forthcoming.

\section{Phase-field model and simulations}
\label{sec:simulations}
For simulating the two-phase flow problem of dynamic electrowetting, we adopt a phase-field (or diffuse-interface) approach.
The interface is described by the order parameter field $\phi$ which attains the values $\pm 1$ respectively in the two phases, and interpolates between the two across the diffuse interface of thickness $\epsilon$.
In the limit $\varepsilon \to 0$, the equations should reproduce the correct sharp-interface physics (see Ref.~\cite{linga2018b}).
A thermodynamically consistent phase-field model fit for our purpose was formulated by \citet{campillo-funollet2012}, and is given by the following set of equations:
\begin{gather}
  \begin{split}
    \pdt (\rho (\phi) \v u) + \div \left( \rho(\phi) \v u \otimes \v u \right) \\
- \div \left[ 2 \mu (\phi) \mathcal D \v u + \v u \otimes \rho'(\phi) M(\phi) \grad g_\phi \right] + \grad p\\
    = - \phi \grad g_\phi - \sum_j {c_j} \grad g_{c_j},
  \end{split} \label{eq:PF_NS1}\\
  \div \v u = 0, \label{eq:PF_NS2}\\
  \pdt \phi + \v u \cdot \grad \phi - \div(M(\phi) \grad g_\phi ) = 0, \label{eq:PF_PF1} \\
  \pdt c_j + \v u \cdot \grad c_j - \div ( D_j (\phi) c_j \grad g_{c_j}) = 0, \label{eq:PF_c} \\
  \div ( \varepsilon (\phi) \grad V ) = - \rho_e. \label{eq:PF_V}
\end{gather}
Here, Eqs.~\eqref{eq:PF_NS1} and \eqref{eq:PF_NS2} are the incompressible Navier--Stokes equations, the Nernst--Planck equation \eqref{eq:PF_c} governs solute transport, and the Poisson equation \eqref{eq:PF_V} determines electrostatic equilibrium.
The phase field $\phi$ takes the value $\phi=-1$ in phase \textbf{s}, and the value $\phi=1$ in phase \textbf{d}.
The (conservative) temporal evolution of $\phi$ is governed by the Cahn--Hilliard equation \eqref{eq:PF_PF1}, wherein the diffusion term is controlled by the phase field mobility $M(\phi)$.
Here, we use the non-linear phase-field mobility
\begin{equation}
  M(\phi) = M_0 (1-\phi^2)_+,
\end{equation}
where $M_0$ is a constant, and $(\cdot)_+ = \max(\cdot, 0)$.

The chemical potential of species $c_\pm$ is given by
\begin{align}
  g_{c_\pm}(c_\pm, \phi) = \ln(c_\pm) + \beta_\pm (\phi) \pm z V,
  \label{eq:PF_chempot}
\end{align}
where $\beta_\pm (\phi)$ is an energy penalty for dissolving ions $c_\pm$ in the phase given by $\phi$.
The chemical potential $g_\phi$ of the phase field $\phi$ is given by:
\begin{multline}
  g_\phi = \frac{3\sigma_{\bf ds}}{2\sqrt{2}} \left[\varepsilon^{-1}W'(\phi) - \varepsilon\laplacian\phi \right] \\
  + \sum_j \beta_j ' (\phi) c_j - \frac 1 2 \epsilon'(\phi) | \grad V |^2 .
  \label{eq:PF_g}
\end{multline}
where $\sigma_{\bf ds}$ is the surface tension, $\epsilon$ is the interface thickness, and $W(\phi)$ is a double well potential.
Here, we adopt the commonly used 
 $W(\phi) = (1-\phi^2)^2/4$.

The density field $\rho$, viscosity field $\mu$, permittivity field $\varepsilon$, solubility energies $\beta_\pm$, and diffusivity fields $D_\pm$ all depend on the phase, i.e.~$\phi$.
In this work, they are given by the following weighted arithmetic averages (WAA):
\begin{align}
  \label{eq:rho_intp}
  \rho (\phi) &= \frac{\rho_{\bf d} + \rho_{\bf s}}{2} + \frac{\rho_{\bf d} - \rho_{\bf s}}{2} \phi, \\
  \label{eq:mu_intp}
  \mu (\phi ) &= \frac{\mu_{\bf d} + \mu_{\bf s}}{2} + \frac{\mu_{\bf d} - \mu_{\bf s}}{2} \phi, \\
  \label{eq:eps_intp}
  \epsilon (\phi) &= \frac{\epsilon_{\bf d} + \epsilon_{\bf s}}{2} + \frac{\epsilon_{\bf d} - \epsilon_{\bf s}}{2} \phi, \\
  \label{eq:Ki_intp}
  D_\pm (\phi) &= \frac{D_{\pm,{\bf d}} + D_{\pm,{\bf s}}}{2} + \frac{D_{\pm,{\bf d}} - D_{\pm,{\bf s}}}{2} \phi, \\
  \beta_\pm (\phi) &= \frac{\beta_{\pm,{\bf d}} + \beta_{\pm,{\bf s}}}{2} + \frac{\beta_{\pm,{\bf d}} - \beta_{\pm,{\bf s}}}{2} \phi.
\end{align}
\citet{tomar2007} found, for a level-set electrohydrodynamics model with smoothed interfacial properties, that using a weighted harmonic average (WHA) for the permittivity yielded more precise results for the electric field than the WAA did.
However, for a model including free charges, \citet{lopez-herrera2011} found no evidence that WHA was superior, and for simplicity we therefore use the WAA for all fields.

\subsection{Boundary conditions}
Most boundary conditions involved in the present work are of Dirichlet type.
We set fixed electric potential at the top and bottom boundaries, and a no-slip condition on the velocity field at the bottom boundary, and fixed concentrations on the top boundary.
Further, we assume a no-flux condition on the concentration fields at the bottom boundary.
With regard to the phase-field, a dynamic wetting boundary condition can be expressed as the following Robin condition \cite{carlson2012}:
\begin{equation}
  \epsilon \tau_w \pdt \phi = \frac{3 \sigma}{2\sqrt{2}} \left[ - \epsilon \hat{\v n} \cdot \grad \phi + \cos(\theta_0) f_w' (\phi) \right],
  \label{eq:BC_phi_dynamic}
\end{equation}
where $\theta_0$ is the prescribed contact angle, $\tau_w$ is a relaxation parameter, and $f_w(\phi) = (2+3\phi-\phi^3)/4$ interpolates smoothly between 0 (at $\phi=-1$) and 1 (at $\phi=1$).
In order not to introduce an additional unknown time scale into the problem, we limit ourselves to considering Eq.~\eqref{eq:BC_phi_dynamic} with $\tau_w = 0$.
Electrowetting with emphasis on contact line pinning was previously studied numerically by \citet{nochetto2014}, who used a generalized Navier boundary condition on the velocity field (cf.~\cite{qian2006}).
However, as contact-line modelling remains phenomenological, we shall leave it for further work.

\subsection{Numerical implementation}
We consider computationally the 2D domain $[0, L_x] \crossprod [0, L_y]$, since as indicated in Fig.~\ref{fig:schematic}, a mirror symmetry is present.
Although alternatively an axially symmetric geometry could have been considered, we consider here the purely two-dimensional case.
In order to mimic a reflective boundary and without loss of generality, we use a free slip condition on the left hand side and a no-flux condition on both electrolyte concentration and electric potential.
The numerical benefits are that this avoids drift of the droplet (due to numerical noise or mesh asymmetries) and limits the computational domain to half the size.

The simulation is initiated with a (\emph{half}) circular droplet cap of area $\pi R_0^2/4$ (in the \emph{half} domain) that forms a contact angle of $\theta_0$ with the surface, and a uniform concentration of both ions is set in the surrounding phase.
At time $t=0$, a potential $V$ is set at the bottom electrode.

To solve the equations numerically we use the finite-element solver \emph{Bernaise} developed by the authors, and presented and validated in a separate work \cite{linga2018b}.
\emph{Bernaise} is written in Python and builds on the FEniCS/Dolfin framework \cite{logg2010,logg2012}.
The solver operates on unstructured meshes and is therefore suitable when different parts of the domain require very different resolutions.

\begin{figure}[htb]
  \includegraphics[width=0.75\columnwidth]{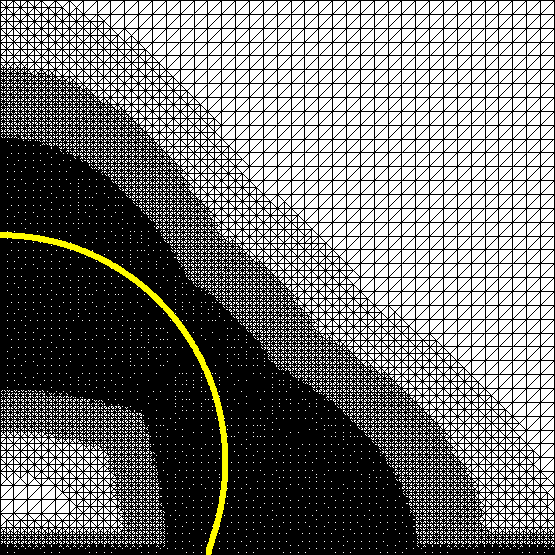}
  \caption{\label{fig:mesh}
    Typical mesh used in simulations.
    The zero-level set of the phase field is shown as a solid yellow line.
  }
\end{figure}

A typical mesh used in the simulations is shown in Fig.~\ref{fig:mesh}.
The mesh is gradually refined near the electrode, to resolve the electrical double layer that arises here.
Further, around the evolving interface, a fine mesh is required; both to resolve the diffuse interface associated with the phase field, and to resolve the Debye layer.
In order to capture the motion of the interface without having to refine adaptively (which is both undesirable for parallelization, and has limited support in FEniCS), the mesh is refined beforehand over an extended area suitable for circle caps with both acute and obtuse contact angles.
For the time integration of the discretized equations, we use the same linear operator splitting scheme as presented in Ref.~\cite{linga2018b}.
With regard to spatial discretization, we use P$_2$ finite elements for the velocity field, and P$_1$ elements for the remaining fields.

\subsection{Physical parameters}
In Sec.~\ref{sec:scaled}, the governing equations were scaled, and since the simulations are carried out in these scaled variables, the results may correspond to a variety of parameter sets.
However, it is interesting to consider concrete physical values in order to relate the numerical experiments to reality.
We consider as an example the components of the ITIES set-up considered by \citet{monroe2006b}, with a nitrobenzene droplet and water surroundings.
The relevant phasic parameters are given in Tab.~\ref{tab:example}.
\begin{table}
  \caption{\label{tab:example} Physical parameters of a water-nitrobenzene system.
  The parameters related to solubility are typical of a monovalent electrolyte such as NaCl.}
  \begin{ruledtabular}
    \begin{tabular}{ c c c c }
      & \multicolumn{2}{c}{Phase $i$} & \\
      Parameter & {\bf d} (nitrobenzene) & {\bf s} (water) & Unit\\
      \hline
      $\epsilon_{r, i}$ & $\simeq 40$ & $\simeq 80$ & -- \\
      $c_i$ & 0 & $0.1$ & M \\
       & 0 & $6\cdot 10^{25}$ & No./m$^3$ \\
      $\lambda_i$ & -- & $\simeq 3$ & nm \\
      $D_{\pm, i}$ & -- & $\simeq 1 \cdot 10^{-9}$ & m$^2$/s \\
      $\rho_i$ & $\simeq 1.2$ g/mL & $\simeq 1.0 \cdot 10^{-3} $ & kg/cm$^3$ \\
      $\mu_i/\rho_i$ & $\simeq 1.7 \cdot 10^{-6} $ m$^2$/s & $\simeq 10^{-6}$ & m$^2$/s \\
    \end{tabular}
  \end{ruledtabular}
\end{table}
Additionally, the surface tension of the water-nitrobenzene interface is (in the order of magnitude) $\sigma_{\bf ds} \simeq 25 \cdot 10^{-3}$ kg/s$^2$ \cite{dupeyrat1969}.
We are now in a position to estimate the expected control parameter $f_0$ defined in Eq.~\eqref{eq:f0}.
Translating back to the dimensional quantities, we have the expression
\begin{equation}
  f_0 = \frac{8 V_T^{3/2} \sqrt{2 z q_e c_0 \epsilon_0 \epsilon_{\rm r,\bf s}}}{\sigma_{\bf ds}},
\end{equation}
which gives a numerical (dimensionless) value of the order $f_0 \simeq 0.3$.
By inspecting \eqref{eq:contact_angle_prediction_ineq}, we see that this value imparts significant deviations from the neutral angle even at moderate $V_T$.
For example, complete dewetting is predicted at $V_0 \simeq 7 V_T $ (assuming the neutral contact angle $\theta_0 = \pi/2$ in the absence of electric field).
For systems with lower surface tension and/or higher concentration, the effect should be stronger.

Inspired by the parameters for the water-nitrobenzene system, we make the simplifying assumptions $\rho_{\bf d} \simeq \rho_{\bf s}$, $\mu_{\bf d} \simeq \mu_{\bf s}$, and $D_- \simeq D_+$.
On the other hand, we choose $\mu_{\bf s}/\rho_{\bf s} \sim D_\pm$ in order to reduce the computation required to equilibrate the charges in the system.
This does not have consequences for the time-asymptotic solution (cf.~Eq.~\eqref{eq:full_G}), and should only have minor consequences for the dynamics.

\section{Results}
\label{sec:results}
Here, we study numerically the dynamic relaxation to an apparent contact angle when an electric field is suddenly turned on.

\subsection{Qualitative description}
\begin{figure}[htb]
  \begin{tikzpicture}
    \draw (0, 0) node[inner sep=0] {
      \includegraphics[width=0.15\textwidth,trim={0 0 3cm 0cm},clip]{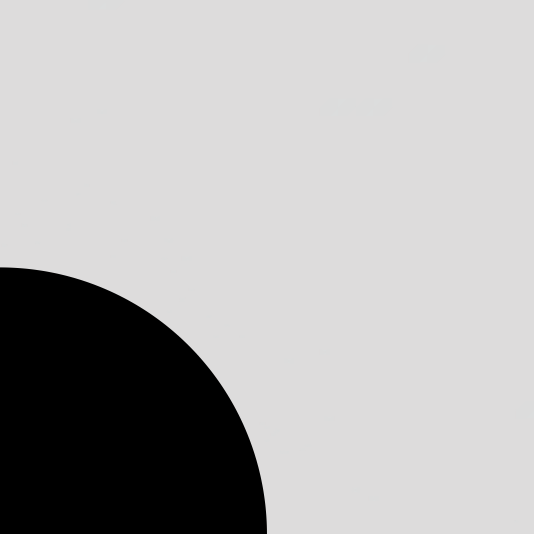}};
    \draw (-1.3,1.35) node [right]{(a) $t=0$};
  \end{tikzpicture}
  \begin{tikzpicture}
    \draw (0, 0) node[inner sep=0] {
      \includegraphics[width=0.15\textwidth,trim={0 0 3cm 0cm},clip]{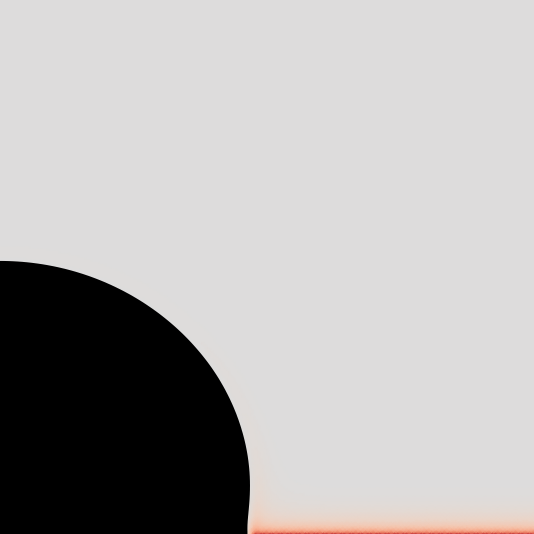}};
    \draw (-1.3,1.35) node [right]{(b) $t=12.5$};
  \end{tikzpicture}
  \begin{tikzpicture}
    \draw (0, 0) node[inner sep=0] {
      \includegraphics[width=0.15\textwidth,trim={0 0 3cm 0cm},clip]{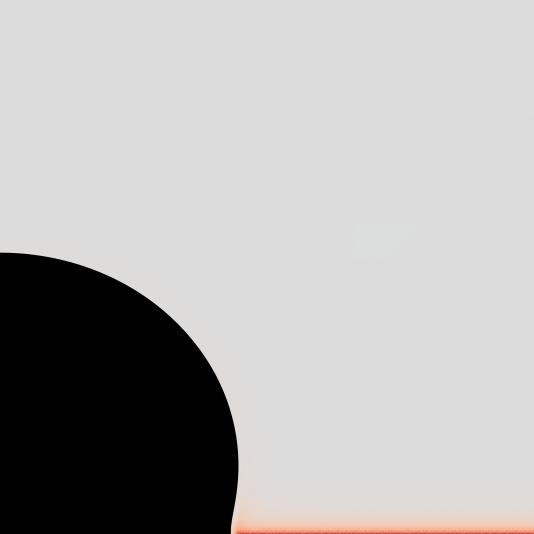}};
    \draw (-1.3,1.35) node [right]{(c) $t=25.0$};
  \end{tikzpicture} \\ \vspace{0.1cm}
  \begin{tikzpicture}
    \draw (0, 0) node[inner sep=0] {
      \includegraphics[width=0.15\textwidth,trim={0 0 3cm 0cm},clip]{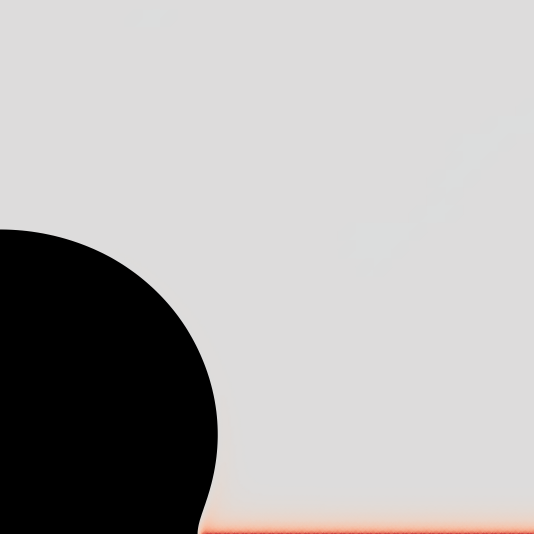}};
    \draw (-1.3,1.35) node [right]{(d) $t=62.5$};
  \end{tikzpicture}
  \begin{tikzpicture}
    \draw (0, 0) node[inner sep=0] {
      \includegraphics[width=0.15\textwidth,trim={0 0 3cm 0cm},clip]{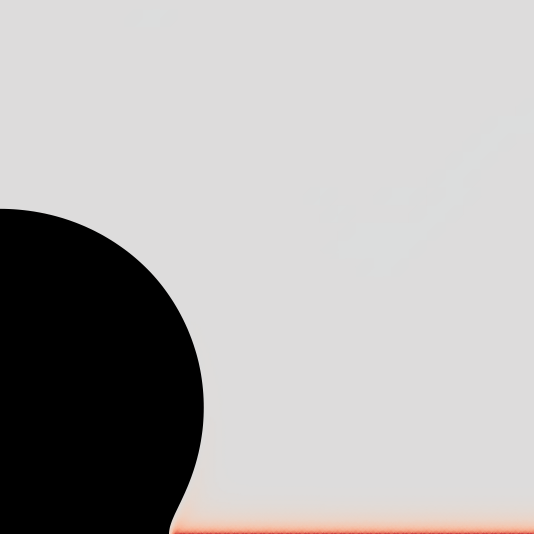}};
    \draw (-1.3,1.35) node [right] {(e) $t=125.0$};
  \end{tikzpicture}
  \begin{tikzpicture}
    \draw (0, 0) node[inner sep=0] {
      \includegraphics[width=0.15\textwidth,trim={0 0 3cm 0cm},clip]{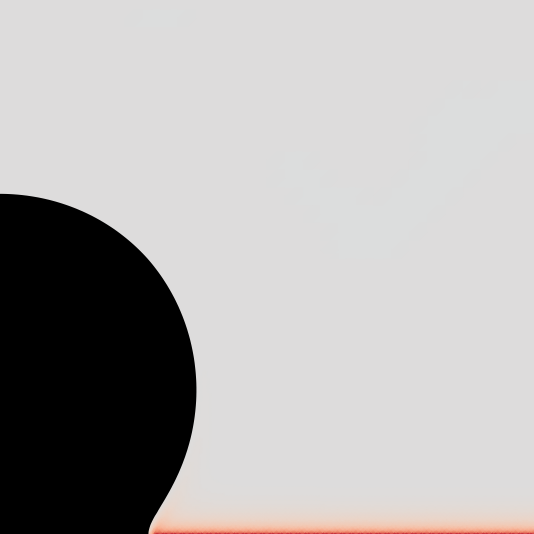}};
    \draw (-1.3,1.35) node [right] {(f) $t=500.0$};
  \end{tikzpicture}
  \caption{\label{fig:electrowetting_anim}
    Relaxation to the apparent contact angle when an electric field is suddenly applied.
    The electric potential difference is turned on to $V=2.5$ at time $t=0$. The red color in the surrounding fluid shows the net charge, and thus represents the EDL.
    (a) to (f) show increasing simulation time.
  }
\end{figure}
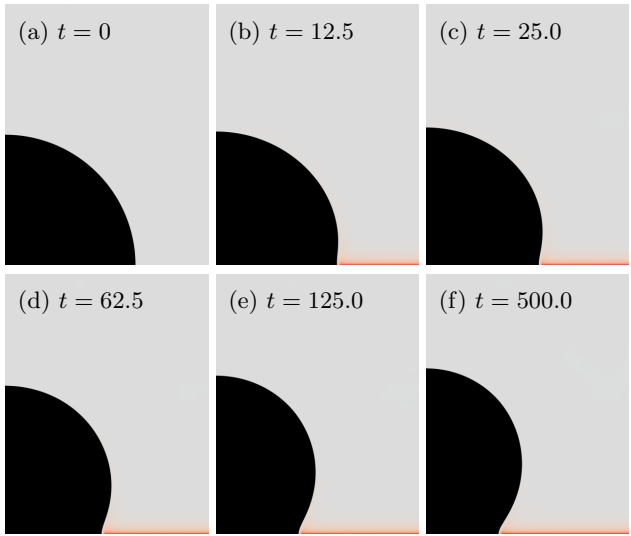

\begin{figure}
  \includegraphics[width=0.9\columnwidth]{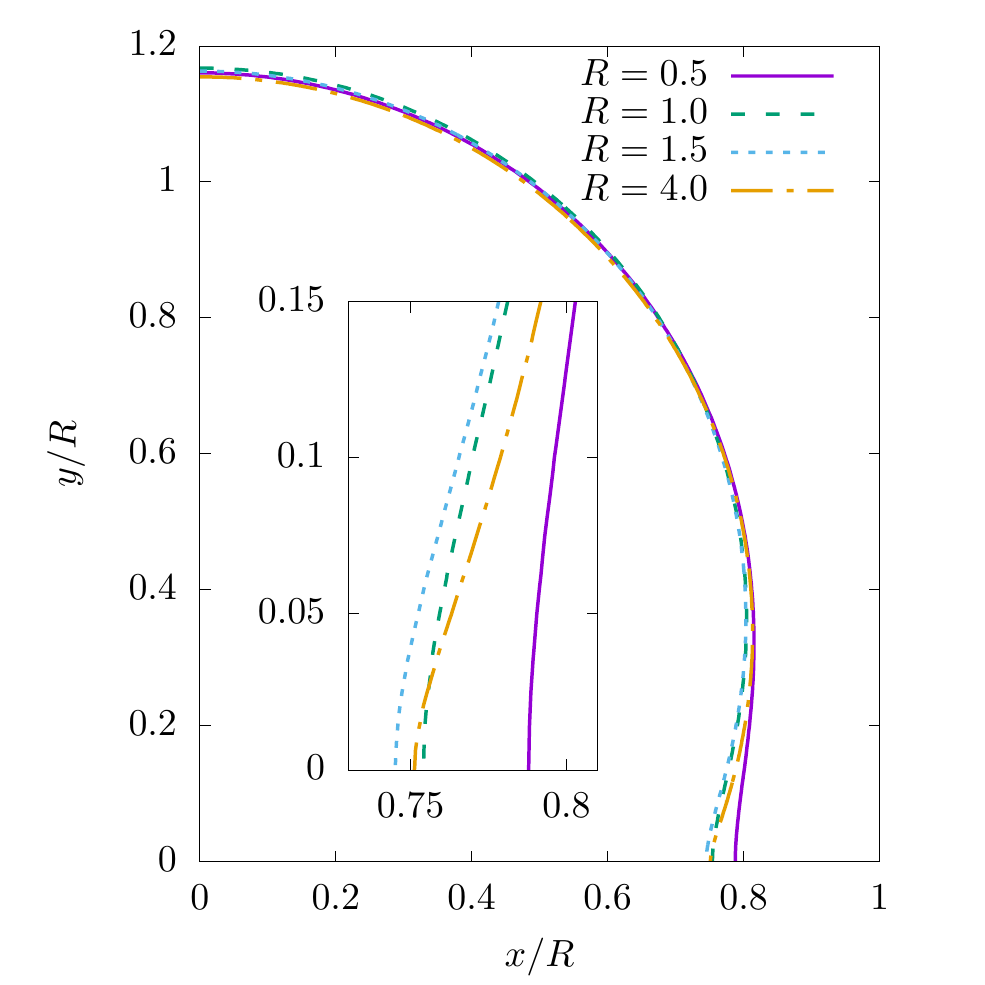}
  \caption{\label{fig:contact_line_comparison} Comparison of the droplet shape for different droplet size, when the Debye length and other parameters are kept constant.
    The Debye length is $\lambda_{\rm s} = \sqrt{\epsilon_{\rm s}/(2 c_0)} \simeq 0.071 $.
    The inset shows a close-up of the contact line.}
\end{figure}

When the potential difference is applied at time $t=0$, charge quickly flows towards the bottom electrode to screen the charge.
Gradually, the contact line moves and an apparent contact angle forms.
In Fig.~\ref{fig:electrowetting_anim} we visualize the relaxation to the apparent contact angle for one specific applied voltage.
Inspecting the local contact angle, we see that the contact angle approaches the strictly enforced angle, here $\theta_0 = \pi/2$.
This is further quantified in Fig.~\ref{fig:contact_line_comparison}, where we compare the final state for the same set-up, same parameters and applied potential, where only the droplet size is varied.
As seen from the figure, the shape of the droplet is fairly robust to the size of the droplet, but is slightly distorted due to the presence of the three-phase contact region.
However, as the Debye length becomes small compared to the droplet radius, the apparent contact angle persists.

\subsection{Contact angle relaxation in time}
\begin{figure}[htb]
  \includegraphics[width=0.99\columnwidth]{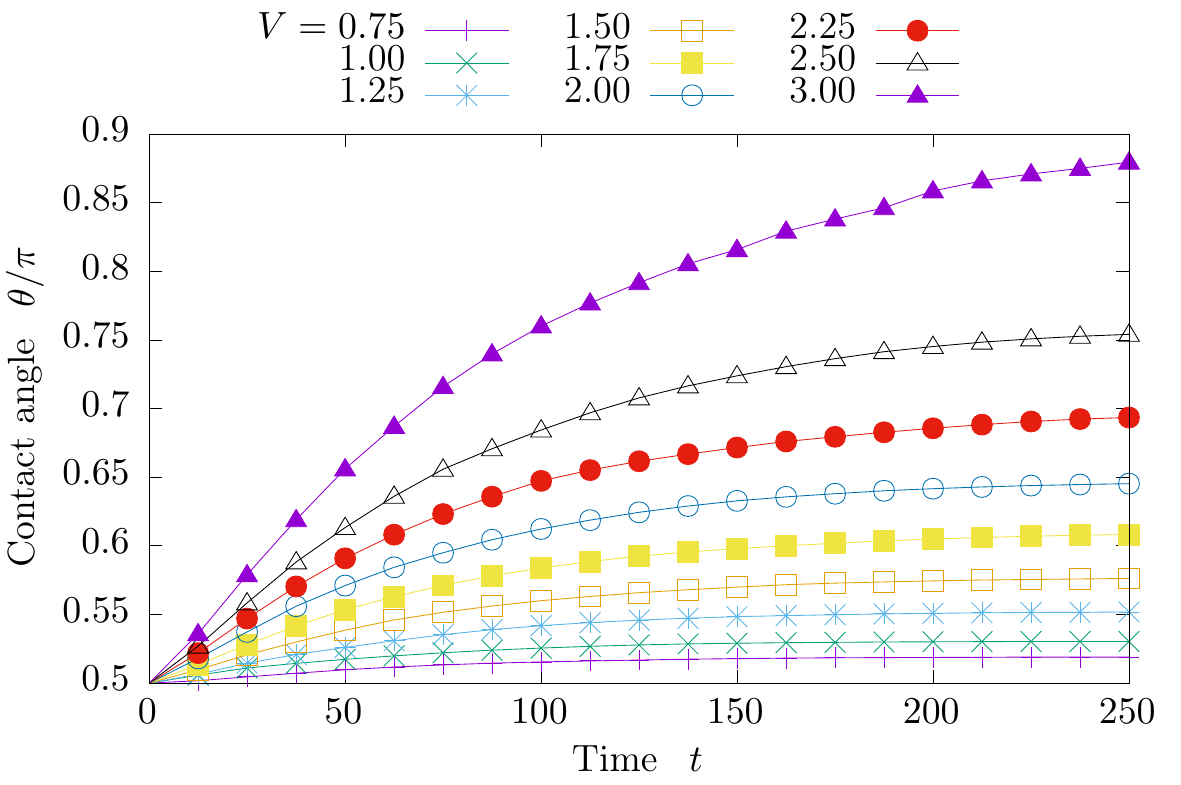}
  \caption{\label{fig:contact_angle_in_time} Contact angle in time for a range of potential drops $V_0$.
  The lines between points are linearly interpolated for visual clarity.}
\end{figure}

We now seek to quantify the evolution of the apparent contact angle through time.
We compute this angle by fitting a semi-circle to the zero-level set of the phase field, for all points where $y \geq 0.1$ ($\sim R_0/10$).
The intersection between this circle and the $y=0$ plane determines the apparent contact angle $\theta$.
In Fig.~\ref{fig:contact_angle_in_time}, we plot the resulting contact angle in time for a range of potential drops.

\subsection{Dependence of the contact angle on applied potential}
\begin{figure}[htb]
  \includegraphics[width=0.9\columnwidth]{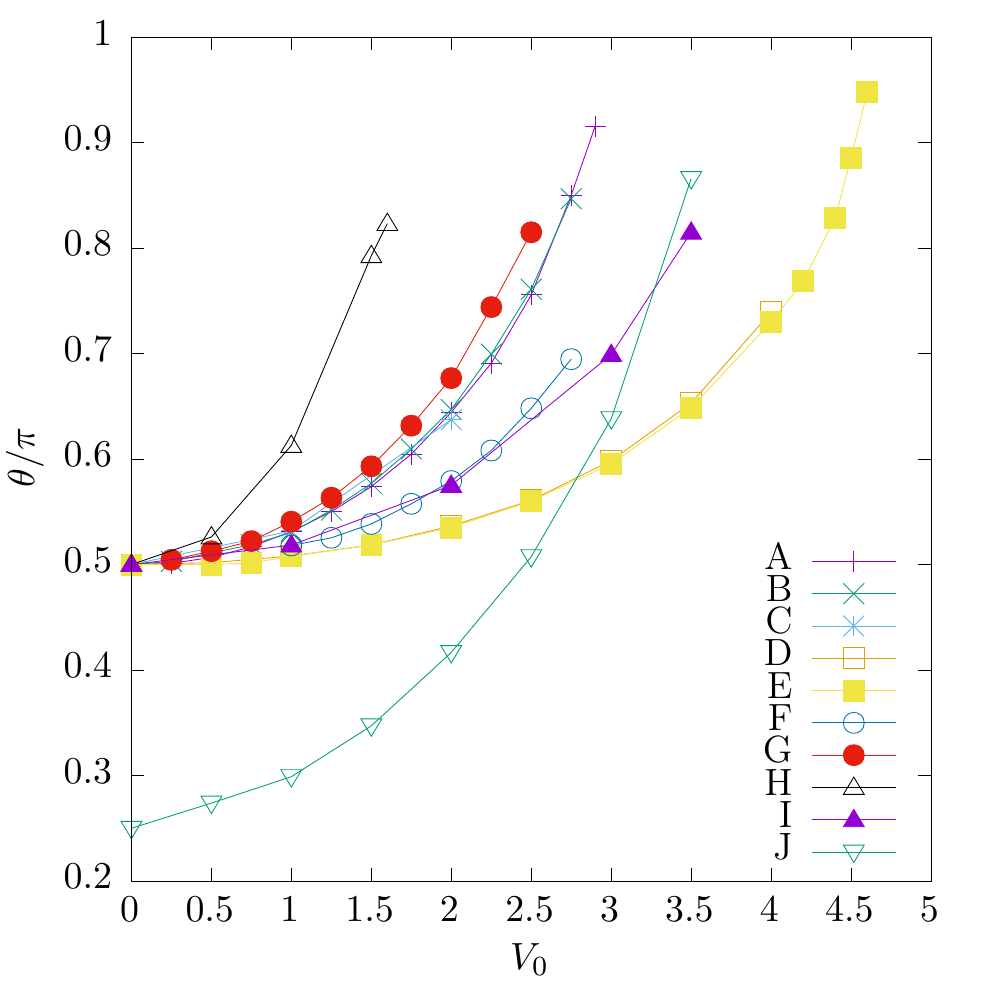}
  \caption{
    \label{fig:contact_angle_of_V_raw}
    We plot the apparent contact angle as a function of applied potential, for a range of parameters.
  The simulation sets A--J correspond to the parameter sets reported in Table \ref{tab:simulations}.}
\end{figure}

In Fig.~\ref{fig:contact_angle_of_V_raw}, we plot the contact angle as a function of applied potential, for a range of different parameter sets.
The parameter sets corresponding to Fig.\ \ref{fig:contact_angle_of_V_raw} are given in Table \ref{tab:simulations}.
The functional form seems to be sensitively dependent on the parameters used.
\begin{table}[htb]
  \caption{\label{tab:simulations} Parameters used in the simulations shown in Fig.\ \ref{fig:contact_angle_of_V_raw}.
    Remaining parameters common for all simulations are $\rho_{\bf d} = \rho_{\bf s} = \mu_{\bf d} = \mu_{\bf s} = 10$, $M_0 = 2 \cdot 10^{-6}$, $D_{\pm,\bf s} = 1$, $D_{\pm,\bf d} = 0.001$, $\beta_{\pm, \bf s} = 0$, $\beta_{\pm, \bf s} = 4$. 
}
  \begin{ruledtabular}
    \begin{tabular}{ c c c c c c c c c }
      Sim. & $R_0$ & $c_0$ & $\epsilon_{\bf s}$ & $\epsilon_{\bf d}$ & $\lambda_{\bf s}$ & $\sigma_{\bf ds}$ & $\theta_0$ & $h_{\rm min} = \varepsilon/2$ \\
      \hline
      A & 1.0 & 10 & 0.1 & 0.2 & 0.071 & 5 & $\pi/2$ & 0.0125 \\
      B & 1.5 & 10 & 0.1 & 0.2 & 0.071 & 5 & $\pi/2$ & 0.0125 \\
      C & 4.0 & 10 & 0.1 & 0.2 & 0.071 & 5 & $\pi/2$ & 0.0125 \\
      D & 1.0 & 1 & 0.1 & 0.2 & 0.22 & 5 & $\pi/2$ & 0.0125 \\
      E & 1.0 & 1 & 0.1 & 0.2 & 0.22 & 5 & $\pi/2$ & 0.025 \\
      F & 1.5 & 10 & 0.1 & 2.0 & 0.071 & 5 & $\pi/2$ & 0.0125 \\
      G & 1.5 & 10 & 0.1 & 0.005 & 0.071 & 5 & $\pi/2$ & 0.0125 \\
      H & 1.5 & 10 & 0.9 & 0.2 & 0.21 & 5 & $\pi/2$ & 0.0125 \\
      I & 1.0 & 10 & 0.1 & 0.2 & 0.071 & 10 & $\pi/2$ & 0.0125 \\
      J & 1.0 & 10 & 0.1 & 0.2 & 0.071 & 5 & $\pi/4$ & 0.0125 \\
    \end{tabular}
  \end{ruledtabular}
\end{table}

\begin{figure}[htb]
  \includegraphics[width=0.99\columnwidth]{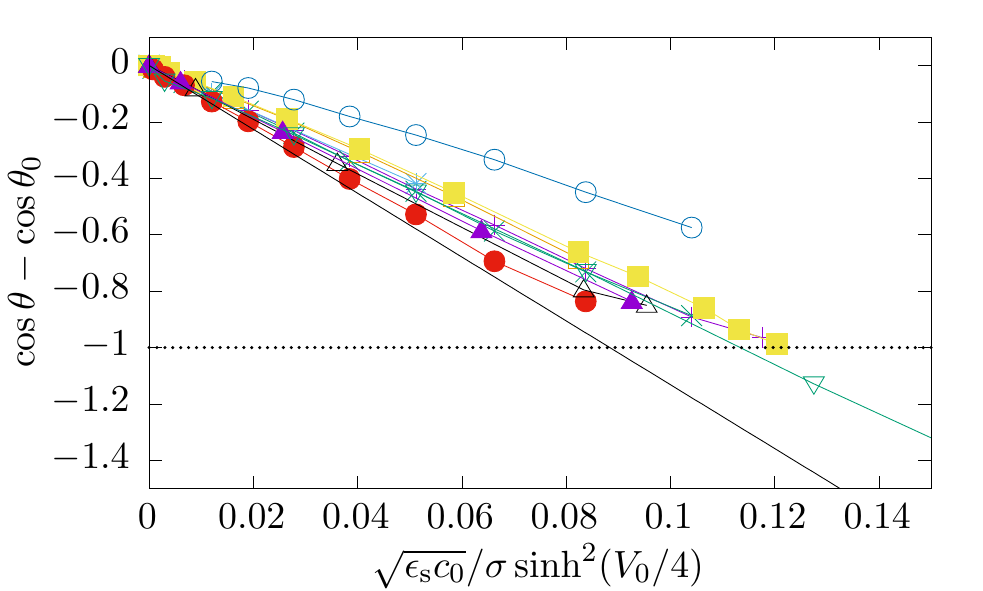}
  \caption{\label{fig:contact_angle_of_V}
    We plot the quantities involved in Eq.~\eqref{eq:contact_angle_prediction} for a range of parameters.
  }
\end{figure}

The prediction of Eq.~\eqref{eq:contact_angle_prediction} suggests that plotting $\cos \theta - \cos \theta_0$ against the composite variable $\left(\sqrt{\epsilon_{\bf s} c_0}/\sigma \right) \sinh^2\left( {V_0}/{4} \right)$, should make the points fall on a straight line, provided that $f$ is independent of $V$.
In Fig.~\ref{fig:contact_angle_of_V}, we show for a range of different parameters the contact angle as a function of this composite variable.
As predicted by Eq.~\eqref{eq:contact_angle_prediction}, it is clear that the proposed functional form matches very well for the entire range until complete dewetting.
Indeed, we find that the points fall onto straight lines for a range of parameters.
Further, the predicted inequality, \eqref{eq:contact_angle_prediction_ineq}, seems to be satisfied for all.
From the figure, it is apparent that the slope of the curves depend mainly on the permittivities in the two phases.

\begin{figure}[htb]
  \includegraphics[width=0.85\columnwidth]{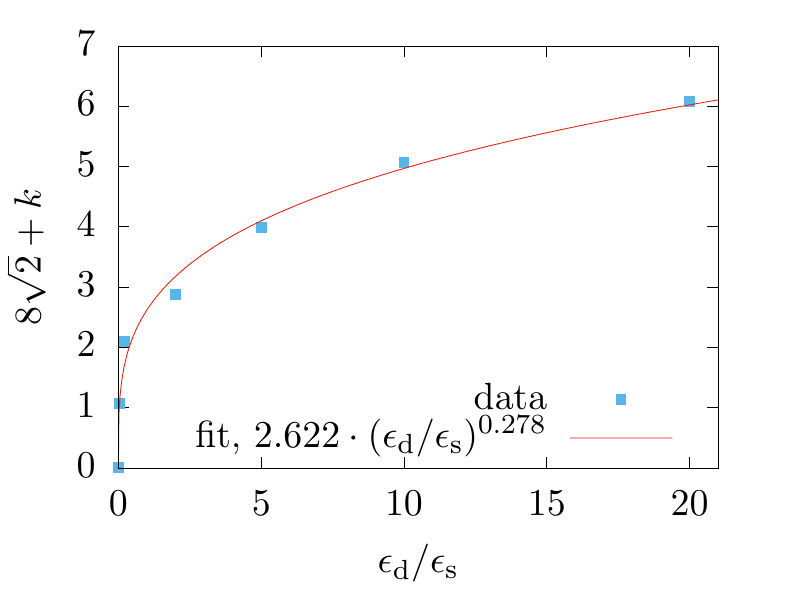}
  \caption{\label{fig:slopes} Computed slopes from the data in Fig.~\ref{fig:contact_angle_of_V} and several other sets of simulations, plotted against permittivity ratio $\epsilon_{\mathrm{d}}/ \epsilon_{\mathrm{s}}$, shown along with a least squares fit.}
\end{figure}

To investigate the role of the permittivities in the two phases, we fit linear slopes to the data plotted in Fig.~\ref{fig:contact_angle_of_V}.
In particular, we use $Y = \cos \theta - \cos \theta_0$ and $X=\left(\sqrt{\epsilon_{\bf s} c_0}/\sigma \right) \sinh^2\left( {V_0}/{4} \right)$, and find for each parameter set the slope $k$ which minimizes the residual of the fit of $Y = k X$ to the $(X, Y)$ data points.
The resulting slopes $k$ (which are all such that $-8\sqrt{2} \leq k \leq 0$)  are plotted in Fig.~\ref{fig:slopes} against the ratio between permittivities $\epsilon_{\mathrm{d}}/ \epsilon_{\mathrm{s}}$ for the respective parameter sets.
\begin{figure}[htb]
  \includegraphics[width=0.85\columnwidth]{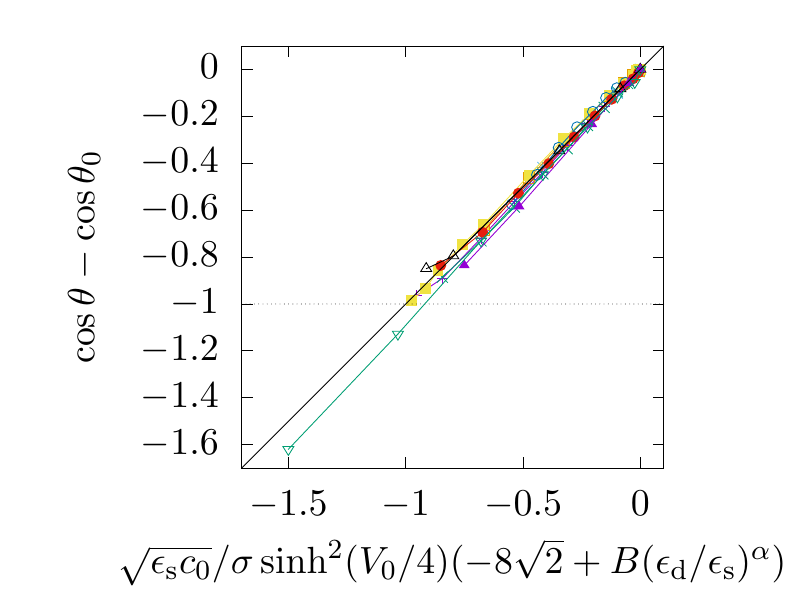}
  \caption{\label{fig:contact_angle_collapse} 
    Collapse of the contact angle data involved in Eq.~\eqref{eq:mainresult}, using the same data as presented in Figs.~\ref{fig:contact_angle_of_V_raw} and \ref{fig:contact_angle_of_V}.
    The black solid line indicates an exact relationship between the ordinate and the abscissa.
  }
\end{figure}
Since we expect $f \leq f_0$, the $y$ axis has been shifted by the numerical prefactor in $f_0$, $8\sqrt{2}$.
We heuristically fit a function $B (\epsilon_{\mathrm{d}}/ \epsilon_{\mathrm{s}})^\alpha$ to these points using least squares, where $B, \alpha$ are (dimensionless) fitting parameters.
The best fit gives $B \simeq 2.6$ and $\alpha=0.28$ (with rather large residuals).
Note that many functional forms would yield fairly equal results.
Our motivation for using exactly this functional form was merely that it required the fewest possible parameters to provide a reasonable fit for the entire range.
Nevertheless, using this scaling function, we are as expected able to collapse the data shown in Figs.~\ref{fig:contact_angle_of_V_raw} and \ref{fig:contact_angle_of_V}.
The resulting relationship is shown in Fig.~\ref{fig:contact_angle_collapse}.
By inspection, moderate deviations from the exact relationship between the abscissa and ordinate quantities can be seen, indicating that improvement could be gained by explicitly taking into account the energy within and around the droplet in the free energy \eqref{eq:DG_1}.
This is, however, out of the scope of the current work.

Within the crude approximations made in deriving \eqref{eq:contact_angle_prediction}, however, the expression
\begin{equation}
 \cos \theta = \cos \theta_0 - \frac{\sqrt{\epsilon_{\mathrm{s}} c_0}}{\sigma} \left[ 8\sqrt{2}-B \left(\frac{\epsilon_{\mathrm{d}}}{\epsilon_{\mathrm{s}}} \right)^{\alpha} \right] \sinh^2(V_0/4)
 \label{eq:mainresult}
\end{equation}
well describes the apparent contact angle for the parameter range considered herein.

\subsection{Relaxation times}
As mentioned previously, it is out of the scope of this work to consider quantitative modelling of the contact line motion.
However, it is in place to inspect the relaxation times associated with the final apparent contact angles presented in the previous subsection.

\begin{figure}[htb]
  \includegraphics[width=0.9\columnwidth]{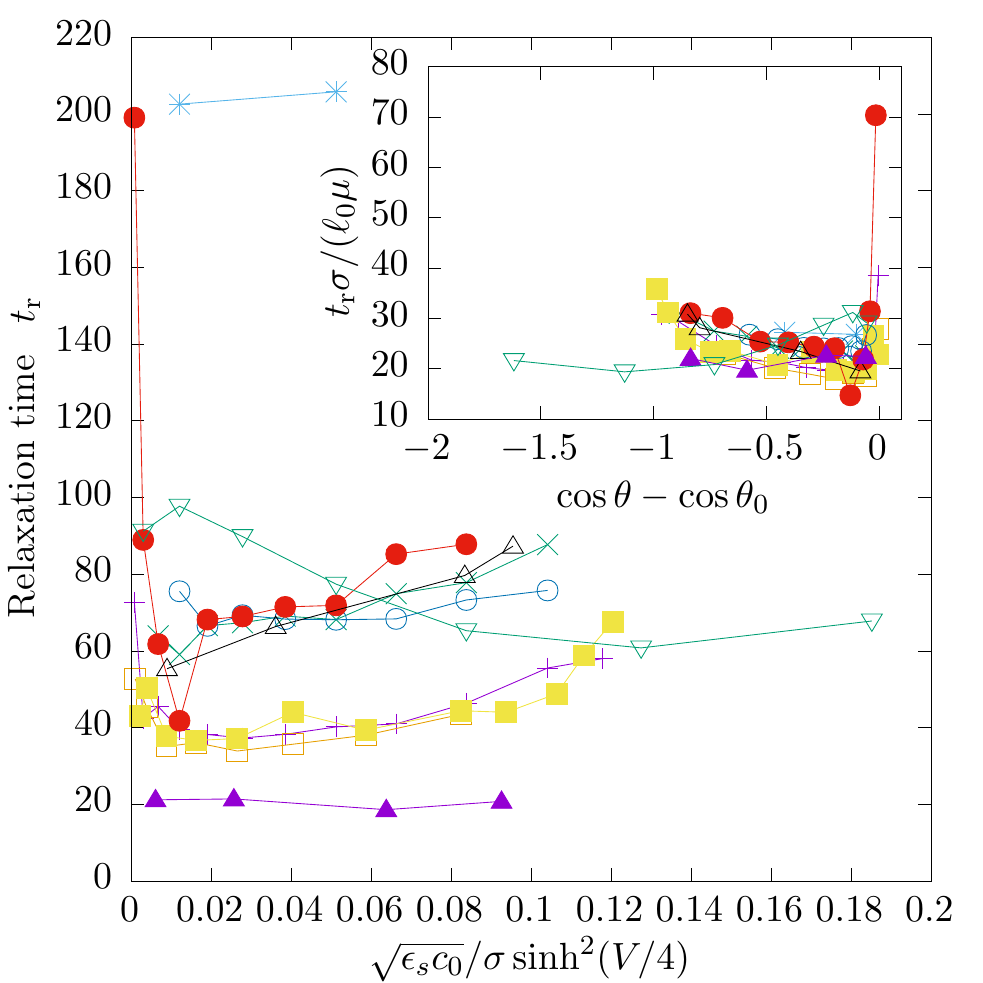}
  \caption{\label{fig:relax} 
    The relaxation times for our simulations obtained by fitting exponential functions to the contact angle in time.
    The data correspond to what is shown in Fig.~\ref{fig:contact_angle_of_V_raw}.
    Inset: Data collapse obtained by using a dimensionless relaxation time based on surface tension $\sigma$, viscosity $\mu$ and the initial wetting length scale $\ell_0$.
    The outliers at low potential/contact angles are due to the poor fit of an exponential function to the data when the contact angle changes only slightly.
  }
\end{figure}

We estimate the relaxation times $t_{\rm r}$ by fitting an exponential function, $C + C' \exp(-t/t_{\rm r})$ to the contact angles as function of time $t$ (cf.~Fig.~\ref{fig:contact_angle_in_time}), where $C, C', t_{\rm r}$ are considered fitting parameters.
In the main panel of Fig.~\ref{fig:relax}, we show the relaxation times that correspond to the final contact angles shown in Fig.~\ref{fig:contact_angle_of_V_raw}.
The relaxation times are fairly constant for each parameter set.
Deviations are noticeable when the applied voltage is low, i.e.~when the contact angle only changes very slightly and the exponential fit becomes unreliable.
Further, at higher $V_0$, the apparent contact angle becomes very obtuse and thus $\theta$ becomes sensitive to the circular fit.
The slight drift seen in the relaxation times should be attributed to that.

On dimensional grounds, we might expect, for flows dominated by viscous and capillary forces,
\begin{equation}
  t_{\rm r} \sim \frac{\ell_0\mu}{\sigma},
\end{equation}
where $\ell_0$ is a typical length scale which we take to be the length scale of the wetted area in the initial state.
This is further admissible since have not introduced any (pinning) dissipation at the moving-contact line in our model.
We check this by plotting the dimensionless quantity $t_{\rm r} {\sigma}/({R\mu})$ against e.g.~the quantity $\cos \theta-\cos\theta_0$, and the resulting plot is shown in the inset of Fig.~\ref{fig:relax}.
The data points collapse fairly well, indicating that the time scale identified above \emph{is} the relevant time scale in our simulations.

\subsection{Comparison to effective modelling}
\begin{figure}[htb]
  \includegraphics[width=0.9\columnwidth]{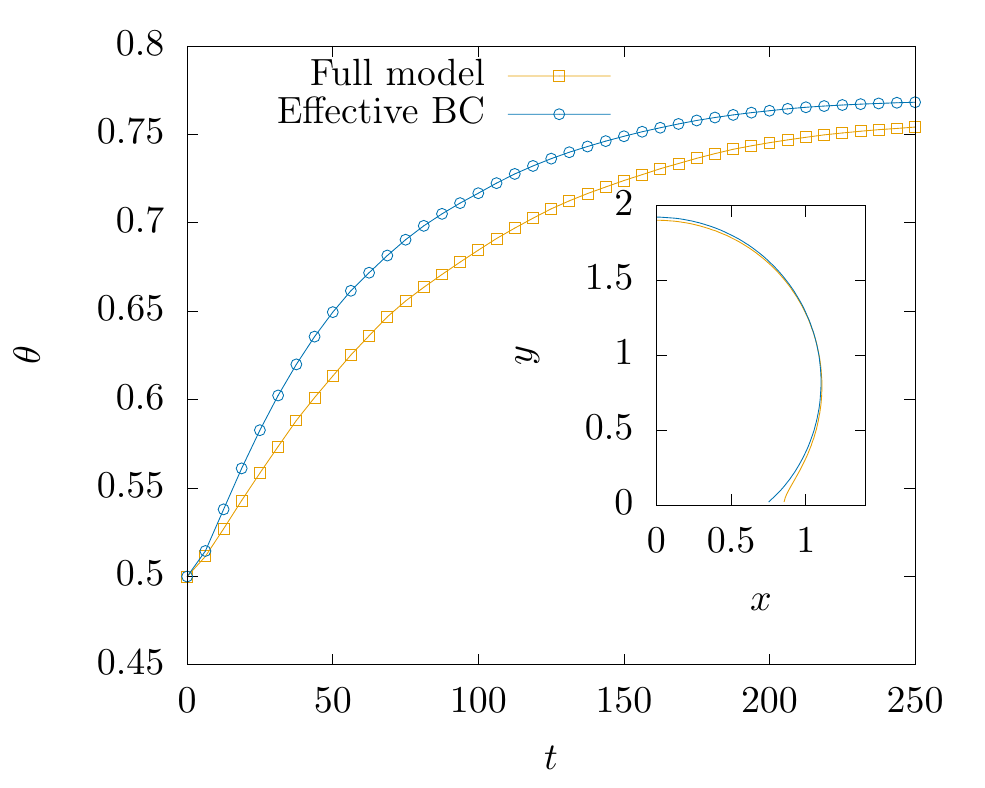}
  \caption{\label{fig:compare_effective}
    Comparison of the evolution of the apparent contact angle as a function of time for the full model, including electrochemistry, and two-phase flow without direct resolution of the electrodynamics but instead using the boundary condition \eqref{eq:effective_BC} and the relationship \eqref{eq:mainresult}.
  }
\end{figure}

As suggested by Eq.~\eqref{eq:mainresult}, it might be useful to avoid simulating dynamic electrowetting using the full model, and instead incorporate the result as a modified contact angle boundary condition.
Recalling Eq.~\eqref{eq:BC_phi_dynamic} (putting again $\tau_w = 0$), we may simply replace $\theta_0$ by the expression for $\theta (V_0)$ given by Eq.~\eqref{eq:mainresult}.
This yields the phase-field boundary condition
\begin{equation}
  \epsilon \hat{\v n} \cdot \grad \phi = \cos \left[ \theta (V_0) \right] f_w' (\phi).
  \label{eq:effective_BC}
\end{equation}
Now, we carry out a direct comparison between the full model and the effective model, where the whole electrochemistry.
In Fig.~\ref{fig:compare_effective} we show a direct comparison of the time evolution (for a simulation at $V_0=2.5$) between these two approaches.
As can be seen from the figure, the \emph{effective BC} approach leads to a faster relaxation to the final contact angle.
The latter also slightly overshoots compared to the simulations using the full model, as can be seen from the inset of Fig.~\ref{fig:compare_effective}.
Hence, the two approaches differ, but not necessarily significantly more than the variations seen within the simulations using the full model, as documented in Figs.~\ref{fig:relax} and \ref{fig:contact_angle_collapse}.
This indicates that the effective BC approach is admissible, but that further modelling might be necessary to quantitatively model the contact line motion.

\section{Discussion}
\label{sec:conclusion}
Compared to similar models of electrowetting \cite{eck2009,nochetto2014,campillo-funollet2012}, we have here used a model that accounts for different ions, and where the conductivity is dependent on the local ion concentration instead of being held fixed.
We have studied systematically the effect of varying the applied potential as well as other physical parameters.
We confirm the results by \citet{mugele2007}, that the contact angle observed by Lippmann is a macroscopic \emph{apparent} contact angle (also commented by e.g.~\cite{eck2009}). 
For a conducting system, the key length scale is the insulator thickness $d$ and the apparent contact angle is only observed on scales beyond $d$. 
In our case, we consider an equipotential boundary, and therefore the length scale that controls the apparent contact angle is the Debye length $\lambda_{\rm s}$ in the surrounding fluid.

\citet{monroe2006} considered theoretically a set-up where ions were dissolved in both phases, whereas we considered the case, where the droplet phase contained no ions.
The Poisson--Boltzmann solution to the electrolyte system presented in Ref.~\cite{monroe2006} was not applicable in our case.
Nonetheless, our numerical experiments have shown that a simplified and slightly heuristically generalized version of the prediction in Ref.~\cite{monroe2006} provides a good description of the contact angle as function of the the applied potential -- even for relatively large Debye lengths.

Clearly, many approximations underpin our results.
First, for the Nernst--Planck equations to hold, we are limited to ideal (i.e.~weak) ionic solutions.
High concentrations would probably not be compatible with the assumption of impermeable interfaces.
Due to resolution requirements, we have limited ourselves to two-dimensional simulations.
Future studies building on the present work should consider axisymmetric or fully three-dimensional geometries.
From a theoretical point of view, a derivation explicitly taking into account e.g.~the energy-minimizing electrostatic potential distribution along the droplet interface, could improve the suggested relation between contact angle and applied potential.

The assumption of a circular droplet geometry (away from the three-phase contact line) may fail when the surface tension, at least compared to the Maxwell stresses, becomes small.
Hence, the results presented are only expected to hold for high surface tension.
We emphasize that although the results presented herein (e.g.~Eq.~\eqref{eq:mainresult}) should clearly not be used outside their domain of validity, the work presented yields a recipe for extending the covered parameter space.
Further, we demonstrated here that numerically resolving electrical double layers constitute an alternative route to obtaining very obtuse or acute contact angles in diffuse-interface simulations of two-phase flow with boundaries.

As mentioned previously, we have not attempted to model contact line friction quantitatively, since this remains in itself an important direction of research.
A next step could be to include the generalized Navier slip boundary condition \cite{qian2003,qian2006}, as was done by \citet{nochetto2014}.
Finally, we did not consider any direct dependency between surface energies and the applied potential.
In general, it would require more detailed modelling to reproduce all the electrochemical effects that are present in experimental settings.

\section{Conclusion}
Controlling wetting properties of two-phase systems is desirable for a wide range of applications.
We have in this paper considered how an applied electric field can control the wetting properties of an electrolytic two-phase system.
To this end, the electrowetting set-up of a droplet sitting on top of an isolated conductor, and where an electrolyte is dissolved in the surrounding phase, was numerically simulated.
This was achieved using a phase-field model for the full electrokinetic two-phase flow problem. We confirmed observations of similar systems from the literature \cite{mugele2007}, i.e.~that an \emph{apparent} contact angle forms on scales beyond the Debye length, which characterizes the extent of the electric double layer.
A main result of our work is summarized in our expression for the effective contact angle, Eq.~\eqref{eq:mainresult}, which was motivated by predictions from nonlinear Poisson--Boltzmann theory.
For models operating on larger scales, the use of such an effective contact angle can greatly improve the computational efficiency. 

\begin{acknowledgments}
This project has received funding from the European Union's Horizon 2020 research and innovation program through Marie Curie initial training networks under grant agreement 642976 (NanoHeal), and from the Villum Foundation through the grant ``Earth Patterns.''
\end{acknowledgments}

\bibliographystyle{apsrev4-1}
\bibliography{references}

 \newcommand{\acis}{Adv. Colloid Interface Sci.}
  \newcommand{\appphyslett}{Appl. Phys. Lett.} \newcommand{\arfm}{Ann. Rev.
  Fluid Mech.} \newcommand{\commcompphys}{Commun. Comput. Phys.}
  \newcommand{\cpc}{Comput. Phys. Commun.} \newcommand{\csurfa}{Colloids Surf.
  A} \newcommand{\ejam}{Eur. J. Appl. Math.} \newcommand{\energyfuels}{Energy
  Fuels} \newcommand{\gji}{Geophys. J. Int.} \newcommand{\ieeetps}{IEEE Trans.
  Plasma Sci.} \newcommand{\ijhmt}{Int. J. Heat Mass Transf.}
  \newcommand{\ijms}{Int. J. Mol. Sci.} \newcommand{\ijmf}{Int. J. Multiph.
  Flow} \newcommand{\ijnmf}{Int. J. Numer. Methods Fluids} \newcommand{\jap}{J.
  Appl. Phys.} \newcommand{\jast}{J. Adhes. Sci. Technol.}
  \newcommand{\jcis}{J. Colloid Interface Sci.} \newcommand{\jcompm}{J. Comp.
  Math.} \newcommand{\jcompp}{J. Comput. Phys.} \newcommand{\jfm}{J. Fluid
  Mech.} \newcommand{\jmems}{J. Microelectromech. Syst.} \newcommand{\jpc}{J.
  Phys. Chem.} \newcommand{\jpcm}{J. Phys. Condens. Matter}
  \newcommand{\labchip}{Lab Chip} \newcommand{\mmmas}{Math. Models Methods
  Appl. Sci.} \newcommand{\natgeo}{Nat. Geosci.} \newcommand{\physfluids}{Phys.
  Fluids} \newcommand{\physreve}{Phys. Rev. E} \newcommand{\physrevlett}{Phys.
  Rev. Lett.} \newcommand{\prsa}{Proc. Royal Soc. A}
  \newcommand{\revmodphys}{Rev. Mod. Phys.} \newcommand{\siap}{SIAM J. Appl.
  Math.}
\begin{thebibliography}{40}%
\makeatletter
\providecommand \@ifxundefined [1]{%
 \@ifx{#1\undefined}
}%
\providecommand \@ifnum [1]{%
 \ifnum #1\expandafter \@firstoftwo
 \else \expandafter \@secondoftwo
 \fi
}%
\providecommand \@ifx [1]{%
 \ifx #1\expandafter \@firstoftwo
 \else \expandafter \@secondoftwo
 \fi
}%
\providecommand \natexlab [1]{#1}%
\providecommand \enquote  [1]{``#1''}%
\providecommand \bibnamefont  [1]{#1}%
\providecommand \bibfnamefont [1]{#1}%
\providecommand \citenamefont [1]{#1}%
\providecommand \href@noop [0]{\@secondoftwo}%
\providecommand \href [0]{\begingroup \@sanitize@url \@href}%
\providecommand \@href[1]{\@@startlink{#1}\@@href}%
\providecommand \@@href[1]{\endgroup#1\@@endlink}%
\providecommand \@sanitize@url [0]{\catcode `\\12\catcode `\$12\catcode
  `\&12\catcode `\#12\catcode `\^12\catcode `\_12\catcode `\%12\relax}%
\providecommand \@@startlink[1]{}%
\providecommand \@@endlink[0]{}%
\providecommand \url  [0]{\begingroup\@sanitize@url \@url }%
\providecommand \@url [1]{\endgroup\@href {#1}{\urlprefix }}%
\providecommand \urlprefix  [0]{URL }%
\providecommand \Eprint [0]{\href }%
\providecommand \doibase [0]{http://dx.doi.org/}%
\providecommand \selectlanguage [0]{\@gobble}%
\providecommand \bibinfo  [0]{\@secondoftwo}%
\providecommand \bibfield  [0]{\@secondoftwo}%
\providecommand \translation [1]{[#1]}%
\providecommand \BibitemOpen [0]{}%
\providecommand \bibitemStop [0]{}%
\providecommand \bibitemNoStop [0]{.\EOS\space}%
\providecommand \EOS [0]{\spacefactor3000\relax}%
\providecommand \BibitemShut  [1]{\csname bibitem#1\endcsname}%
\let\auto@bib@innerbib\@empty
\bibitem [{\citenamefont {Pollack}\ \emph {et~al.}(2002)\citenamefont
  {Pollack}, \citenamefont {Shenderov},\ and\ \citenamefont
  {Fair}}]{pollack2002}%
  \BibitemOpen
  \bibfield  {author} {\bibinfo {author} {\bibfnamefont {M.}~\bibnamefont
  {Pollack}}, \bibinfo {author} {\bibfnamefont {A.}~\bibnamefont {Shenderov}},
  \ and\ \bibinfo {author} {\bibfnamefont {R.}~\bibnamefont {Fair}},\ }\href
  {\doibase 10.1039/b110474h} {\bibfield  {journal} {\bibinfo  {journal}
  {\labchip}\ }\textbf {\bibinfo {volume} {2}},\ \bibinfo {pages} {96}
  (\bibinfo {year} {2002})}\BibitemShut {NoStop}%
\bibitem [{\citenamefont {Srinivasan}\ \emph {et~al.}(2004)\citenamefont
  {Srinivasan}, \citenamefont {Pamula},\ and\ \citenamefont
  {Fair}}]{srinivasan2004}%
  \BibitemOpen
  \bibfield  {author} {\bibinfo {author} {\bibfnamefont {V.}~\bibnamefont
  {Srinivasan}}, \bibinfo {author} {\bibfnamefont {V.~K.}\ \bibnamefont
  {Pamula}}, \ and\ \bibinfo {author} {\bibfnamefont {R.~B.}\ \bibnamefont
  {Fair}},\ }\href {\doibase 10.1039/b403341h} {\bibfield  {journal} {\bibinfo
  {journal} {\labchip}\ }\textbf {\bibinfo {volume} {4}},\ \bibinfo {pages}
  {310} (\bibinfo {year} {2004})}\BibitemShut {NoStop}%
\bibitem [{\citenamefont {Lee}\ and\ \citenamefont {Kim}(2000)}]{lee2000}%
  \BibitemOpen
  \bibfield  {author} {\bibinfo {author} {\bibfnamefont {J.}~\bibnamefont
  {Lee}}\ and\ \bibinfo {author} {\bibfnamefont {C.-J.}\ \bibnamefont {Kim}},\
  }\href {\doibase 10.1109/84.846697} {\bibfield  {journal} {\bibinfo
  {journal} {\jmems}\ }\textbf {\bibinfo {volume} {9}},\ \bibinfo {pages} {171}
  (\bibinfo {year} {2000})}\BibitemShut {NoStop}%
\bibitem [{\citenamefont {Beni}\ and\ \citenamefont
  {Hackwood}(1981)}]{beni1981}%
  \BibitemOpen
  \bibfield  {author} {\bibinfo {author} {\bibfnamefont {G.}~\bibnamefont
  {Beni}}\ and\ \bibinfo {author} {\bibfnamefont {S.}~\bibnamefont
  {Hackwood}},\ }\href {\doibase 10.1063/1.92322} {\bibfield  {journal}
  {\bibinfo  {journal} {\appphyslett}\ }\textbf {\bibinfo {volume} {38}},\
  \bibinfo {pages} {207} (\bibinfo {year} {1981})}\BibitemShut {NoStop}%
\bibitem [{\citenamefont {Beni}\ and\ \citenamefont {Tenan}(1981)}]{beni1981b}%
  \BibitemOpen
  \bibfield  {author} {\bibinfo {author} {\bibfnamefont {G.}~\bibnamefont
  {Beni}}\ and\ \bibinfo {author} {\bibfnamefont {M.}~\bibnamefont {Tenan}},\
  }\href {\doibase 10.1063/1.329822} {\bibfield  {journal} {\bibinfo  {journal}
  {\jap}\ }\textbf {\bibinfo {volume} {52}},\ \bibinfo {pages} {6011} (\bibinfo
  {year} {1981})}\BibitemShut {NoStop}%
\bibitem [{\citenamefont {Beni}\ \emph {et~al.}(1982)\citenamefont {Beni},
  \citenamefont {Hackwood},\ and\ \citenamefont {Jackel}}]{beni1982}%
  \BibitemOpen
  \bibfield  {author} {\bibinfo {author} {\bibfnamefont {G.}~\bibnamefont
  {Beni}}, \bibinfo {author} {\bibfnamefont {S.}~\bibnamefont {Hackwood}}, \
  and\ \bibinfo {author} {\bibfnamefont {J.}~\bibnamefont {Jackel}},\ }\href
  {\doibase 10.1063/1.92920} {\bibfield  {journal} {\bibinfo  {journal}
  {\appphyslett}\ }\textbf {\bibinfo {volume} {40}},\ \bibinfo {pages} {912}
  (\bibinfo {year} {1982})}\BibitemShut {NoStop}%
\bibitem [{\citenamefont {Hayes}\ and\ \citenamefont
  {Feenstra}(2003)}]{hayes2003}%
  \BibitemOpen
  \bibfield  {author} {\bibinfo {author} {\bibfnamefont {R.~A.}\ \bibnamefont
  {Hayes}}\ and\ \bibinfo {author} {\bibfnamefont {B.~J.}\ \bibnamefont
  {Feenstra}},\ }\href {\doibase 10.1038/nature01988} {\bibfield  {journal}
  {\bibinfo  {journal} {Nature}\ }\textbf {\bibinfo {volume} {425}},\ \bibinfo
  {pages} {383} (\bibinfo {year} {2003})}\BibitemShut {NoStop}%
\bibitem [{\citenamefont {Hassenkam}\ \emph {et~al.}(2011)\citenamefont
  {Hassenkam}, \citenamefont {Pedersen}, \citenamefont {Dalby}, \citenamefont
  {Austad},\ and\ \citenamefont {Stipp}}]{hassenkam2011}%
  \BibitemOpen
  \bibfield  {author} {\bibinfo {author} {\bibfnamefont {T.}~\bibnamefont
  {Hassenkam}}, \bibinfo {author} {\bibfnamefont {C.~S.}\ \bibnamefont
  {Pedersen}}, \bibinfo {author} {\bibfnamefont {K.}~\bibnamefont {Dalby}},
  \bibinfo {author} {\bibfnamefont {T.}~\bibnamefont {Austad}}, \ and\ \bibinfo
  {author} {\bibfnamefont {S.~L.~S.}\ \bibnamefont {Stipp}},\ }\href {\doibase
  10.1016/j.colsurfa.2011.09.025} {\bibfield  {journal} {\bibinfo  {journal}
  {\csurfa}\ }\textbf {\bibinfo {volume} {390}},\ \bibinfo {pages} {179}
  (\bibinfo {year} {2011})}\BibitemShut {NoStop}%
\bibitem [{\citenamefont {Hilner}\ \emph {et~al.}(2015)\citenamefont {Hilner},
  \citenamefont {Andersson}, \citenamefont {Hassenkam}, \citenamefont
  {Matthiesen}, \citenamefont {Salino},\ and\ \citenamefont
  {Stipp}}]{hilner2015}%
  \BibitemOpen
  \bibfield  {author} {\bibinfo {author} {\bibfnamefont {E.}~\bibnamefont
  {Hilner}}, \bibinfo {author} {\bibfnamefont {M.~P.}\ \bibnamefont
  {Andersson}}, \bibinfo {author} {\bibfnamefont {T.}~\bibnamefont
  {Hassenkam}}, \bibinfo {author} {\bibfnamefont {J.}~\bibnamefont
  {Matthiesen}}, \bibinfo {author} {\bibfnamefont {P.}~\bibnamefont {Salino}},
  \ and\ \bibinfo {author} {\bibfnamefont {S.~L.~S.}\ \bibnamefont {Stipp}},\
  }\href {\doibase 10.1038/srep09933} {\bibfield  {journal} {\bibinfo
  {journal} {Scientific reports}\ }\textbf {\bibinfo {volume} {5}},\ \bibinfo
  {pages} {9933} (\bibinfo {year} {2015})}\BibitemShut {NoStop}%
\bibitem [{\citenamefont {RezaeiDoust}\ \emph {et~al.}(2009)\citenamefont
  {RezaeiDoust}, \citenamefont {Puntervold}, \citenamefont {Strand},\ and\
  \citenamefont {Austad}}]{rezaeidoust2009}%
  \BibitemOpen
  \bibfield  {author} {\bibinfo {author} {\bibfnamefont {A.}~\bibnamefont
  {RezaeiDoust}}, \bibinfo {author} {\bibfnamefont {T.}~\bibnamefont
  {Puntervold}}, \bibinfo {author} {\bibfnamefont {S.}~\bibnamefont {Strand}},
  \ and\ \bibinfo {author} {\bibfnamefont {T.}~\bibnamefont {Austad}},\ }\href
  {\doibase 10.1021/ef900185q} {\bibfield  {journal} {\bibinfo  {journal}
  {\energyfuels}\ }\textbf {\bibinfo {volume} {23}},\ \bibinfo {pages} {4479}
  (\bibinfo {year} {2009})}\BibitemShut {NoStop}%
\bibitem [{\citenamefont {Pedersen}\ \emph {et~al.}(2016)\citenamefont
  {Pedersen}, \citenamefont {Hassenkam}, \citenamefont {Ceccato}, \citenamefont
  {Dalby}, \citenamefont {Mogensen},\ and\ \citenamefont
  {Stipp}}]{pedersen2016}%
  \BibitemOpen
  \bibfield  {author} {\bibinfo {author} {\bibfnamefont {N.}~\bibnamefont
  {Pedersen}}, \bibinfo {author} {\bibfnamefont {T.}~\bibnamefont {Hassenkam}},
  \bibinfo {author} {\bibfnamefont {M.}~\bibnamefont {Ceccato}}, \bibinfo
  {author} {\bibfnamefont {K.~N.}\ \bibnamefont {Dalby}}, \bibinfo {author}
  {\bibfnamefont {K.}~\bibnamefont {Mogensen}}, \ and\ \bibinfo {author}
  {\bibfnamefont {S.~L.~S.}\ \bibnamefont {Stipp}},\ }\href {\doibase
  10.1021/acs.energyfuels.5b02562} {\bibfield  {journal} {\bibinfo  {journal}
  {\energyfuels}\ }\textbf {\bibinfo {volume} {30}},\ \bibinfo {pages} {3768}
  (\bibinfo {year} {2016})}\BibitemShut {NoStop}%
\bibitem [{\citenamefont {Hiorth}\ \emph {et~al.}(2010)\citenamefont {Hiorth},
  \citenamefont {Cathles},\ and\ \citenamefont {Madland}}]{hiorth2010}%
  \BibitemOpen
  \bibfield  {author} {\bibinfo {author} {\bibfnamefont {A.}~\bibnamefont
  {Hiorth}}, \bibinfo {author} {\bibfnamefont {L.}~\bibnamefont {Cathles}}, \
  and\ \bibinfo {author} {\bibfnamefont {M.}~\bibnamefont {Madland}},\ }\href
  {\doibase 10.1007/s11242-010-9543-6} {\bibfield  {journal} {\bibinfo
  {journal} {Transport in porous media}\ }\textbf {\bibinfo {volume} {85}},\
  \bibinfo {pages} {1} (\bibinfo {year} {2010})}\BibitemShut {NoStop}%
\bibitem [{\citenamefont {Lippmann}(1875)}]{lippmann1875}%
  \BibitemOpen
  \bibfield  {author} {\bibinfo {author} {\bibfnamefont {G.}~\bibnamefont
  {Lippmann}},\ }\emph {\bibinfo {title} {Relations entre les
  ph{\'e}nom{\`e}nes {\'e}lectriques et capillaires}},\ \href@noop {} {Ph.D.
  thesis},\ \bibinfo  {school} {Sorbonne} (\bibinfo {year} {1875})\BibitemShut
  {NoStop}%
\bibitem [{\citenamefont {Mugele}\ and\ \citenamefont
  {Baret}(2005)}]{mugele2005}%
  \BibitemOpen
  \bibfield  {author} {\bibinfo {author} {\bibfnamefont {F.}~\bibnamefont
  {Mugele}}\ and\ \bibinfo {author} {\bibfnamefont {J.-C.}\ \bibnamefont
  {Baret}},\ }\href {\doibase 10.1088/0953-8984/17/28/R01} {\bibfield
  {journal} {\bibinfo  {journal} {\jpcm}\ }\textbf {\bibinfo {volume} {17}},\
  \bibinfo {pages} {R705} (\bibinfo {year} {2005})}\BibitemShut {NoStop}%
\bibitem [{\citenamefont {Monroe}\ \emph
  {et~al.}(2006{\natexlab{a}})\citenamefont {Monroe}, \citenamefont {Daikhin},
  \citenamefont {Urbakh},\ and\ \citenamefont {Kornyshev}}]{monroe2006}%
  \BibitemOpen
  \bibfield  {author} {\bibinfo {author} {\bibfnamefont {C.~W.}\ \bibnamefont
  {Monroe}}, \bibinfo {author} {\bibfnamefont {L.~I.}\ \bibnamefont {Daikhin}},
  \bibinfo {author} {\bibfnamefont {M.}~\bibnamefont {Urbakh}}, \ and\ \bibinfo
  {author} {\bibfnamefont {A.~A.}\ \bibnamefont {Kornyshev}},\ }\href {\doibase
  10.1103/PhysRevLett.97.136102} {\bibfield  {journal} {\bibinfo  {journal}
  {\physrevlett}\ }\textbf {\bibinfo {volume} {97}},\ \bibinfo {pages} {136102}
  (\bibinfo {year} {2006}{\natexlab{a}})}\BibitemShut {NoStop}%
\bibitem [{\citenamefont {Levich}(1962)}]{levich1962}%
  \BibitemOpen
  \bibfield  {author} {\bibinfo {author} {\bibfnamefont {V.~G.}\ \bibnamefont
  {Levich}},\ }\href@noop {} {\emph {\bibinfo {title} {Physicochemical
  hydrodynamics}}}\ (\bibinfo  {publisher} {Prentice Hall},\ \bibinfo {year}
  {1962})\BibitemShut {NoStop}%
\bibitem [{\citenamefont {Mugele}(2009)}]{mugele2009b}%
  \BibitemOpen
  \bibfield  {author} {\bibinfo {author} {\bibfnamefont {F.}~\bibnamefont
  {Mugele}},\ }\href {\doibase 10.1039/B904493K} {\bibfield  {journal}
  {\bibinfo  {journal} {Soft Matter}\ }\textbf {\bibinfo {volume} {5}},\
  \bibinfo {pages} {3377} (\bibinfo {year} {2009})}\BibitemShut {NoStop}%
\bibitem [{\citenamefont {Mugele}\ and\ \citenamefont
  {Buehrle}(2007)}]{mugele2007}%
  \BibitemOpen
  \bibfield  {author} {\bibinfo {author} {\bibfnamefont {F.}~\bibnamefont
  {Mugele}}\ and\ \bibinfo {author} {\bibfnamefont {J.}~\bibnamefont
  {Buehrle}},\ }\href {\doibase 10.1088/0953-8984/19/37/375112} {\bibfield
  {journal} {\bibinfo  {journal} {\jpcm}\ }\textbf {\bibinfo {volume} {19}},\
  \bibinfo {pages} {375112} (\bibinfo {year} {2007})}\BibitemShut {NoStop}%
\bibitem [{\citenamefont {Eck}\ \emph {et~al.}(2009)\citenamefont {Eck},
  \citenamefont {Fontelos}, \citenamefont {Gr{\"u}n}, \citenamefont
  {Klingbeil},\ and\ \citenamefont {Vantzos}}]{eck2009}%
  \BibitemOpen
  \bibfield  {author} {\bibinfo {author} {\bibfnamefont {C.}~\bibnamefont
  {Eck}}, \bibinfo {author} {\bibfnamefont {M.}~\bibnamefont {Fontelos}},
  \bibinfo {author} {\bibfnamefont {G.}~\bibnamefont {Gr{\"u}n}}, \bibinfo
  {author} {\bibfnamefont {F.}~\bibnamefont {Klingbeil}}, \ and\ \bibinfo
  {author} {\bibfnamefont {O.}~\bibnamefont {Vantzos}},\ }\href {\doibase
  10.4171/IFB/211} {\bibfield  {journal} {\bibinfo  {journal} {Interfaces and
  free boundaries}\ }\textbf {\bibinfo {volume} {11}},\ \bibinfo {pages} {259}
  (\bibinfo {year} {2009})}\BibitemShut {NoStop}%
\bibitem [{\citenamefont {Monroe}\ \emph
  {et~al.}(2006{\natexlab{b}})\citenamefont {Monroe}, \citenamefont {Daikhin},
  \citenamefont {Urbakh},\ and\ \citenamefont {Kornyshev}}]{monroe2006b}%
  \BibitemOpen
  \bibfield  {author} {\bibinfo {author} {\bibfnamefont {C.~W.}\ \bibnamefont
  {Monroe}}, \bibinfo {author} {\bibfnamefont {L.~I.}\ \bibnamefont {Daikhin}},
  \bibinfo {author} {\bibfnamefont {M.}~\bibnamefont {Urbakh}}, \ and\ \bibinfo
  {author} {\bibfnamefont {A.~A.}\ \bibnamefont {Kornyshev}},\ }\href {\doibase
  10.1088/0953-8984/18/10/009} {\bibfield  {journal} {\bibinfo  {journal}
  {\jpcm}\ }\textbf {\bibinfo {volume} {18}},\ \bibinfo {pages} {2837}
  (\bibinfo {year} {2006}{\natexlab{b}})}\BibitemShut {NoStop}%
\bibitem [{\citenamefont {Taylor}(1966)}]{taylor1966}%
  \BibitemOpen
  \bibfield  {author} {\bibinfo {author} {\bibfnamefont {G.}~\bibnamefont
  {Taylor}},\ }\href@noop {} {\bibfield  {journal} {\bibinfo  {journal}
  {Proceedings of the Royal Society of London. Series A. Mathematical and
  Physical Sciences}\ }\textbf {\bibinfo {volume} {291}},\ \bibinfo {pages}
  {159} (\bibinfo {year} {1966})}\BibitemShut {NoStop}%
\bibitem [{\citenamefont {Melcher}\ and\ \citenamefont
  {Taylor}(1969)}]{melcher1969}%
  \BibitemOpen
  \bibfield  {author} {\bibinfo {author} {\bibfnamefont {J.}~\bibnamefont
  {Melcher}}\ and\ \bibinfo {author} {\bibfnamefont {G.}~\bibnamefont
  {Taylor}},\ }\href {\doibase 10.1146/annurev.fl.01.010169.000551} {\bibfield
  {journal} {\bibinfo  {journal} {\arfm}\ }\textbf {\bibinfo {volume} {1}},\
  \bibinfo {pages} {111} (\bibinfo {year} {1969})}\BibitemShut {NoStop}%
\bibitem [{\citenamefont {Schnitzer}\ and\ \citenamefont
  {Yariv}(2015)}]{schnitzer2015}%
  \BibitemOpen
  \bibfield  {author} {\bibinfo {author} {\bibfnamefont {O.}~\bibnamefont
  {Schnitzer}}\ and\ \bibinfo {author} {\bibfnamefont {E.}~\bibnamefont
  {Yariv}},\ }\href {\doibase 10.1017/jfm.2015.242} {\bibfield  {journal}
  {\bibinfo  {journal} {Journal of Fluid Mechanics}\ }\textbf {\bibinfo
  {volume} {773}},\ \bibinfo {pages} {1} (\bibinfo {year} {2015})}\BibitemShut
  {NoStop}%
\bibitem [{\citenamefont {Zholkovskij}\ \emph {et~al.}(2002)\citenamefont
  {Zholkovskij}, \citenamefont {Masliyah},\ and\ \citenamefont
  {Czarnecki}}]{zholkovskij2002}%
  \BibitemOpen
  \bibfield  {author} {\bibinfo {author} {\bibfnamefont {E.~K.}\ \bibnamefont
  {Zholkovskij}}, \bibinfo {author} {\bibfnamefont {J.~H.}\ \bibnamefont
  {Masliyah}}, \ and\ \bibinfo {author} {\bibfnamefont {J.}~\bibnamefont
  {Czarnecki}},\ }\href {\doibase 10.1017/S0022112002001441} {\bibfield
  {journal} {\bibinfo  {journal} {Journal of Fluid Mechanics}\ }\textbf
  {\bibinfo {volume} {472}},\ \bibinfo {pages} {1} (\bibinfo {year}
  {2002})}\BibitemShut {NoStop}%
\bibitem [{\citenamefont {Saville}(1997)}]{saville1997}%
  \BibitemOpen
  \bibfield  {author} {\bibinfo {author} {\bibfnamefont {D.~A.}\ \bibnamefont
  {Saville}},\ }\href {\doibase 10.1146/annurev.fluid.29.1.27} {\bibfield
  {journal} {\bibinfo  {journal} {\arfm}\ }\textbf {\bibinfo {volume} {29}},\
  \bibinfo {pages} {27} (\bibinfo {year} {1997})}\BibitemShut {NoStop}%
\bibitem [{\citenamefont {Berry}\ \emph {et~al.}(2013)\citenamefont {Berry},
  \citenamefont {Davidson},\ and\ \citenamefont {Harvie}}]{berry2013}%
  \BibitemOpen
  \bibfield  {author} {\bibinfo {author} {\bibfnamefont {J.}~\bibnamefont
  {Berry}}, \bibinfo {author} {\bibfnamefont {M.}~\bibnamefont {Davidson}}, \
  and\ \bibinfo {author} {\bibfnamefont {D.~J.}\ \bibnamefont {Harvie}},\
  }\href {\doibase 10.1016/j.jcp.2013.05.026} {\bibfield  {journal} {\bibinfo
  {journal} {Journal of Computational Physics}\ }\textbf {\bibinfo {volume}
  {251}},\ \bibinfo {pages} {209} (\bibinfo {year} {2013})}\BibitemShut
  {NoStop}%
\bibitem [{\citenamefont {Tomar}\ \emph {et~al.}(2007)\citenamefont {Tomar},
  \citenamefont {Gerlach}, \citenamefont {Biswas}, \citenamefont {Alleborn},
  \citenamefont {Sharma}, \citenamefont {Durst}, \citenamefont {Welch},\ and\
  \citenamefont {Delgado}}]{tomar2007}%
  \BibitemOpen
  \bibfield  {author} {\bibinfo {author} {\bibfnamefont {G.}~\bibnamefont
  {Tomar}}, \bibinfo {author} {\bibfnamefont {D.}~\bibnamefont {Gerlach}},
  \bibinfo {author} {\bibfnamefont {G.}~\bibnamefont {Biswas}}, \bibinfo
  {author} {\bibfnamefont {N.}~\bibnamefont {Alleborn}}, \bibinfo {author}
  {\bibfnamefont {A.}~\bibnamefont {Sharma}}, \bibinfo {author} {\bibfnamefont
  {F.}~\bibnamefont {Durst}}, \bibinfo {author} {\bibfnamefont
  {S.}~\bibnamefont {Welch}}, \ and\ \bibinfo {author} {\bibfnamefont
  {A.}~\bibnamefont {Delgado}},\ }\href {\doibase 10.1016/j.jcp.2007.09.003}
  {\bibfield  {journal} {\bibinfo  {journal} {\jcompp}\ }\textbf {\bibinfo
  {volume} {227}},\ \bibinfo {pages} {1267 } (\bibinfo {year}
  {2007})}\BibitemShut {NoStop}%
\bibitem [{\citenamefont {L{\'o}pez-Herrera}\ \emph {et~al.}(2011)\citenamefont
  {L{\'o}pez-Herrera}, \citenamefont {Popinet},\ and\ \citenamefont
  {Herrada}}]{lopez-herrera2011}%
  \BibitemOpen
  \bibfield  {author} {\bibinfo {author} {\bibfnamefont {J.}~\bibnamefont
  {L{\'o}pez-Herrera}}, \bibinfo {author} {\bibfnamefont {S.}~\bibnamefont
  {Popinet}}, \ and\ \bibinfo {author} {\bibfnamefont {M.}~\bibnamefont
  {Herrada}},\ }\href {\doibase 10.1016/j.jcp.2010.11.042} {\bibfield
  {journal} {\bibinfo  {journal} {Journal of Computational Physics}\ }\textbf
  {\bibinfo {volume} {230}},\ \bibinfo {pages} {1939} (\bibinfo {year}
  {2011})}\BibitemShut {NoStop}%
\bibitem [{\citenamefont {Nochetto}\ \emph {et~al.}(2014)\citenamefont
  {Nochetto}, \citenamefont {Salgado},\ and\ \citenamefont
  {Walker}}]{nochetto2014}%
  \BibitemOpen
  \bibfield  {author} {\bibinfo {author} {\bibfnamefont {R.~H.}\ \bibnamefont
  {Nochetto}}, \bibinfo {author} {\bibfnamefont {A.~J.}\ \bibnamefont
  {Salgado}}, \ and\ \bibinfo {author} {\bibfnamefont {S.~W.}\ \bibnamefont
  {Walker}},\ }\href {\doibase 10.1142/S0218202513500474} {\bibfield  {journal}
  {\bibinfo  {journal} {\mmmas}\ }\textbf {\bibinfo {volume} {24}},\ \bibinfo
  {pages} {67} (\bibinfo {year} {2014})}\BibitemShut {NoStop}%
\bibitem [{\citenamefont {Lu}\ \emph {et~al.}(2007)\citenamefont {Lu},
  \citenamefont {Glasner}, \citenamefont {Bertozzi},\ and\ \citenamefont
  {Kim}}]{lu2007}%
  \BibitemOpen
  \bibfield  {author} {\bibinfo {author} {\bibfnamefont {H.-W.}\ \bibnamefont
  {Lu}}, \bibinfo {author} {\bibfnamefont {K.}~\bibnamefont {Glasner}},
  \bibinfo {author} {\bibfnamefont {A.}~\bibnamefont {Bertozzi}}, \ and\
  \bibinfo {author} {\bibfnamefont {C.-J.}\ \bibnamefont {Kim}},\ }\href
  {\doibase 10.1017/S0022112007008154} {\bibfield  {journal} {\bibinfo
  {journal} {\jfm}\ }\textbf {\bibinfo {volume} {590}},\ \bibinfo {pages} {411}
  (\bibinfo {year} {2007})}\BibitemShut {NoStop}%
\bibitem [{\citenamefont {Walker}\ and\ \citenamefont
  {Shapiro}(2006)}]{walker2006}%
  \BibitemOpen
  \bibfield  {author} {\bibinfo {author} {\bibfnamefont {S.~W.}\ \bibnamefont
  {Walker}}\ and\ \bibinfo {author} {\bibfnamefont {B.}~\bibnamefont
  {Shapiro}},\ }\href {\doibase 10.1109/JMEMS.2006.878876} {\bibfield
  {journal} {\bibinfo  {journal} {\jmems}\ }\textbf {\bibinfo {volume} {15}},\
  \bibinfo {pages} {986} (\bibinfo {year} {2006})}\BibitemShut {NoStop}%
\bibitem [{\citenamefont {Walker}\ \emph {et~al.}(2009)\citenamefont {Walker},
  \citenamefont {Shapiro},\ and\ \citenamefont {Nochetto}}]{walker2009}%
  \BibitemOpen
  \bibfield  {author} {\bibinfo {author} {\bibfnamefont {S.~W.}\ \bibnamefont
  {Walker}}, \bibinfo {author} {\bibfnamefont {B.}~\bibnamefont {Shapiro}}, \
  and\ \bibinfo {author} {\bibfnamefont {R.~H.}\ \bibnamefont {Nochetto}},\
  }\href {\doibase 10.1063/1.3254022} {\bibfield  {journal} {\bibinfo
  {journal} {\physfluids}\ }\textbf {\bibinfo {volume} {21}},\ \bibinfo {pages}
  {102103} (\bibinfo {year} {2009})}\BibitemShut {NoStop}%
\bibitem [{\citenamefont {Campillo-Funollet}\ \emph {et~al.}(2012)\citenamefont
  {Campillo-Funollet}, \citenamefont {Gr{\"u}n},\ and\ \citenamefont
  {Klingbeil}}]{campillo-funollet2012}%
  \BibitemOpen
  \bibfield  {author} {\bibinfo {author} {\bibfnamefont {E.}~\bibnamefont
  {Campillo-Funollet}}, \bibinfo {author} {\bibfnamefont {G.}~\bibnamefont
  {Gr{\"u}n}}, \ and\ \bibinfo {author} {\bibfnamefont {F.}~\bibnamefont
  {Klingbeil}},\ }\href {\doibase 10.1137/120861333} {\bibfield  {journal}
  {\bibinfo  {journal} {\siap}\ }\textbf {\bibinfo {volume} {72}},\ \bibinfo
  {pages} {1899} (\bibinfo {year} {2012})}\BibitemShut {NoStop}%
\bibitem [{\citenamefont {Linga}\ \emph {et~al.}(2018)\citenamefont {Linga},
  \citenamefont {Bolet},\ and\ \citenamefont {Mathiesen}}]{linga2018b}%
  \BibitemOpen
  \bibfield  {author} {\bibinfo {author} {\bibfnamefont {G.}~\bibnamefont
  {Linga}}, \bibinfo {author} {\bibfnamefont {A.}~\bibnamefont {Bolet}}, \ and\
  \bibinfo {author} {\bibfnamefont {J.}~\bibnamefont {Mathiesen}},\ }\href@noop
  {} {\enquote {\bibinfo {title} {Bernaise: A flexible framework for two-phase
  electrohydrodynamic flow},}\ } (\bibinfo {year} {2018}),\ \bibinfo {note}
  {submitted}\BibitemShut {NoStop}%
\bibitem [{\citenamefont {Carlson}\ \emph {et~al.}(2012)\citenamefont
  {Carlson}, \citenamefont {Bellani},\ and\ \citenamefont
  {Amberg}}]{carlson2012}%
  \BibitemOpen
  \bibfield  {author} {\bibinfo {author} {\bibfnamefont {A.}~\bibnamefont
  {Carlson}}, \bibinfo {author} {\bibfnamefont {G.}~\bibnamefont {Bellani}}, \
  and\ \bibinfo {author} {\bibfnamefont {G.}~\bibnamefont {Amberg}},\ }\href
  {\doibase 10.1103/PhysRevE.85.045302} {\bibfield  {journal} {\bibinfo
  {journal} {\physreve}\ }\textbf {\bibinfo {volume} {85}},\ \bibinfo {pages}
  {045302} (\bibinfo {year} {2012})}\BibitemShut {NoStop}%
\bibitem [{\citenamefont {Qian}\ \emph {et~al.}(2006)\citenamefont {Qian},
  \citenamefont {Wang},\ and\ \citenamefont {Sheng}}]{qian2006}%
  \BibitemOpen
  \bibfield  {author} {\bibinfo {author} {\bibfnamefont {T.}~\bibnamefont
  {Qian}}, \bibinfo {author} {\bibfnamefont {X.-P.}\ \bibnamefont {Wang}}, \
  and\ \bibinfo {author} {\bibfnamefont {P.}~\bibnamefont {Sheng}},\ }\href
  {\doibase 10.1017/S0022112006001935} {\bibfield  {journal} {\bibinfo
  {journal} {Journal of Fluid Mechanics}\ }\textbf {\bibinfo {volume} {564}},\
  \bibinfo {pages} {333} (\bibinfo {year} {2006})}\BibitemShut {NoStop}%
\bibitem [{\citenamefont {Logg}\ and\ \citenamefont {Wells}(2010)}]{logg2010}%
  \BibitemOpen
  \bibfield  {author} {\bibinfo {author} {\bibfnamefont {A.}~\bibnamefont
  {Logg}}\ and\ \bibinfo {author} {\bibfnamefont {G.~N.}\ \bibnamefont
  {Wells}},\ }\href {\doibase 10.1145/1731022.1731030} {\bibfield  {journal}
  {\bibinfo  {journal} {ACM Trans. Math. Softw.}\ }\textbf {\bibinfo {volume}
  {37}},\ \bibinfo {pages} {20:1} (\bibinfo {year} {2010})}\BibitemShut
  {NoStop}%
\bibitem [{\citenamefont {Logg}\ \emph {et~al.}(2012)\citenamefont {Logg},
  \citenamefont {Mardal},\ and\ \citenamefont {Wells}}]{logg2012}%
  \BibitemOpen
  \bibfield  {author} {\bibinfo {author} {\bibfnamefont {A.}~\bibnamefont
  {Logg}}, \bibinfo {author} {\bibfnamefont {K.-A.}\ \bibnamefont {Mardal}}, \
  and\ \bibinfo {author} {\bibfnamefont {G.}~\bibnamefont {Wells}},\
  }\href@noop {} {\emph {\bibinfo {title} {Automated solution of differential
  equations by the finite element method: The FEniCS book}}},\ Vol.~\bibinfo
  {volume} {84}\ (\bibinfo  {publisher} {Springer Science \& Business Media},\
  \bibinfo {year} {2012})\BibitemShut {NoStop}%
\bibitem [{\citenamefont {Dupeyrat}\ and\ \citenamefont
  {Michel}(1969)}]{dupeyrat1969}%
  \BibitemOpen
  \bibfield  {author} {\bibinfo {author} {\bibfnamefont {M.}~\bibnamefont
  {Dupeyrat}}\ and\ \bibinfo {author} {\bibfnamefont {J.}~\bibnamefont
  {Michel}},\ }\href {\doibase 10.1016/0021-9797(69)90210-0} {\bibfield
  {journal} {\bibinfo  {journal} {Journal of Colloid and Interface Science}\
  }\textbf {\bibinfo {volume} {29}},\ \bibinfo {pages} {605} (\bibinfo {year}
  {1969})}\BibitemShut {NoStop}%
\bibitem [{\citenamefont {Qian}\ \emph {et~al.}(2003)\citenamefont {Qian},
  \citenamefont {Wang},\ and\ \citenamefont {Sheng}}]{qian2003}%
  \BibitemOpen
  \bibfield  {author} {\bibinfo {author} {\bibfnamefont {T.}~\bibnamefont
  {Qian}}, \bibinfo {author} {\bibfnamefont {X.-P.}\ \bibnamefont {Wang}}, \
  and\ \bibinfo {author} {\bibfnamefont {P.}~\bibnamefont {Sheng}},\ }\href
  {\doibase 10.1103/PhysRevE.68.016306} {\bibfield  {journal} {\bibinfo
  {journal} {Physical Review E}\ }\textbf {\bibinfo {volume} {68}},\ \bibinfo
  {pages} {016306} (\bibinfo {year} {2003})}\BibitemShut {NoStop}%
\end{thebibliography}%

\end{document}